\documentclass[aps,pre,amsfonts,superscriptaddress,amssymb,epsfig,graphics,showpacs,11pt]{revtex4}

\usepackage{graphicx}
\usepackage{epsfig}

\newcommand \be {\begin{equation}} 
\newcommand \ee {\end{equation}}

\newcommand \bea {\begin{eqnarray}} 
\newcommand \eea {\end{eqnarray}}

\begin{document}
\newcommand{\<}{\left\langle}
\renewcommand{\>}{\right\rangle}
\newcommand{\ua}{\uparrow}
\newcommand{\da}{\downarrow}
\newcommand{\la}{\leftarrow}
\newcommand{\ra}{\rightarrow}

\def\dblone{\hbox{$1\hskip -4.5pt\vrule depth 0pt height 1.3ex width 1pt
            \vrule depth 0pt height 1.3pt width 0.12em$}}
\def\ex{\text{e}}
\def\uo{\underline{\omega}}
\def\vp{\vec{p}}
\def\vq{\vec{q}}
\def\vI{\vec{I}}
\def\vae{\varepsilon}

\date{\today}

\title{Glassy phases in  Random Heteropolymers with correlated sequences}
\author{M.\ M\"uller$^{\,1}$, M. M\'ezard}
\affiliation{
Laboratoire de Physique Th\'eorique et Mod\`eles Statistiques,\\
Universit\'e Paris-Sud, b\^atiment 100, F-91405 Orsay,
France.}

\author{A. Montanari}
\affiliation{Laboratoire de Physique Th\'{e}orique de l'Ecole Normale
  Sup\'{e}rieure, Paris, France.\footnote{UMR 8549, Unit{\'e} Mixte de Recherche du 
  Centre National de la Recherche Scientifique et de 
  l' Ecole Normale Sup{\'e}rieure.}
}


\begin{abstract} 
We develop a new analytic approach for the study of lattice heteropolymers,
and apply it to copolymers with correlated Markovian sequences.
According to our analysis, heteropolymers present three different dense
phases depending upon the temperature, the nature of
the monomer interactions, and the sequence correlations:
$(i)$ a liquid  phase, $(ii)$ a ``soft glass'' phase,
and $(iii)$ a ``frozen glass'' phase. 
The presence of the new intermediate ``soft glass'' phase is 
predicted for instance in the case of polyampholytes with
sequences that favor the alternation of monomers.

Our approach is based on the cavity method, a refined Bethe Peierls
approximation adapted to frustrated systems. It amounts to a mean field
treatment in which the nearest neighbor correlations, which are crucial in
the dense phases of heteropolymers, are handled exactly. This approach is
powerful and versatile, it can be improved systematically and generalized to
other polymeric systems.
\end{abstract}

\maketitle

\section{Introduction}
In the last 20 years much effort has been devoted to the theoretical
study of heteropolymers \cite{Heteropolymerreview, SfatosShakhnovich97}. One of the main motivations was to
understand the statistical physics of protein folding \cite{PolymerREM,WolynesREM,BryngelsonReview95,Proteinfoldingbook2,Onuchicreview,Grosberg00}.
Despite the insight that has been accumulated, the goal remains distant. 
On the one hand, most analytical studies have been limited
to random bond models \cite{GarelOrland88,ShakhnovichGutin89c} (in which the interaction energies
of all the couples of monomers along the chain are independent random 
variables), or to uncorrelated random copolymer sequences \cite{GarelLeibler94,SfatosGutin93}. However,
there are many indications that sequence correlations induced by natural selection 
play an important role for the folding and stability of proteins.
On the other hand, in this difficult problem,
analytic computations have to resort to some approximations 
which are not easy to control. It is thus important to have a variety of 
different techniques at hand in order to crosscheck the predictions.

In this paper we develop a new tool for
the analytical study of heteropolymers, based on the cavity method
as used in various frustrated systems (a short account of our results
has appeared in \cite{MontanariMullerMezard03a}). We use this method to investigate 
the phase diagram of copolymers with Markovian sequences.
Within our approach we find  copolymers to exist in three distinct dense 
phases (apart from the diluted coil phase at high temperature) depending
upon the structure of the interaction energy matrix, 
the sequence correlations and the temperature: $(i)$ The liquid globule phase
in which distinct monomers are essentially uncorrelated 
and can freely rearrange 
within the globule (apart from obvious constraints on monomers that are close along the chain); $(ii)$ the ``frozen glass'' phase in which the
polymer is stuck in one out of a few well-separated 
low-energy conformations; $(iii)$ a ``soft glass'' phase with broken ergodicity (in the thermodynamic limit) in which
the  thermodynamically relevant conformations 
 form a continuum in configuration space.
This last phase has never been predicted in an analytical computation
(although such a possibility has been envisioned in 
phenomenological models~\cite{PolymerGREM1,Grosberg00}, and a very similar phase seems to be present in the numerical results of \cite{CopolymerPhasedia} on the dynamics of heteropolymers.). 
Albeit frustrated, it has a 
much larger entropy, and appears already at a smaller density
than the usual ``frozen glass'' phase.

Some of the most successful tools used so far in the study of
random heteropolymers are mean field approaches based on the replica method \cite{GarelOrland88, ShakhnovichGutin89c,SfatosGutin93}. 
Crucial to these calculations
was the identification of some relevant  order parameter,
and the proposition of a suitable Ansatz describing the
phase transition in a coupled space of real space coordinates and replica 
indices. This type of approach is potentially very powerful,
but it becomes quite complex for heteropolymers. On the one hand, it requires
a physical intuition for identifying the relevant degrees 
of freedom and of their behavior. On the other hand, 
an Ansatz tailored to describe a certain type of physics 
may hide other, unexpected features.

Our cavity method consists in a refined version of
the Bethe Peierls approximation. While this also represents a 
kind of mean-field approximation, it differs fundamentally from the previous 
ones. Applying the Bethe-Peierls approximation
to lattice heteropolymers allows  to describe self-consistently the 
frustration on a local microscopic level. This approach
 can be thought of as the first step 
in the series of cluster variational (or Kikuchi) approximations \cite{Kikuchi51}.
Its general philosophy consists in keeping track of local
correlations inside some small region exactly, while treating the external
degrees of freedom as an environment whose statistical properties 
have to be determined self-consistently. In the Bethe approximation, the 
only correlations which are treated exactly are the ones between 
neighboring sites on the lattice. This is an
improvement with respect to the na\"{\i}ve mean field that treats 
distinct sites as statistically independent. Moreover, it is the first
of such approximations to be meaningful for polymers, since the backbone
structure induces  strong
correlations between neighbors \cite{Nagle74,AguileraKikuchi1,AguileraKikuchi2,AguileraKikuchi3,AguileraKikuchi4,AguileraKikuchi5}.
Another potential advantage of the cavity method is that
it can be used for one given polymer, without the need to average over
an ensemble of sequences as in the replica method. While in the present work
we  focus on ensemble-averaged properties, one should keep in mind this
possibility which
 could lead to interesting algorithmic developments in
the future.
Finally, the refined Bethe-Peierls approximation is supposed to be exact 
on locally tree-like structures (e.g., on random graphs). 
This is an important feature: It allows one to set up the mean-field analysis 
in a mathematically well-defined way, and its predictions
can be checked against numerical simulations on those 
random ``mean-field'' lattices for which the theory is expected to be exact.

Within our cavity method, any heteropolymer is found to undergo a glass
transition at large enough densities. Two main schemes of
glass transitions can occur, depending on the details of the
sequence, each of them being associated with one of the types of glasses mentioned
above. 

The transition to the frozen glass phase is a discontinuous transition, which
is called random first order, or one step replica symmetry breaking (1RSB)
transition in the replica language. It corresponds to the type of transition
which has been found in many previous studies, of which the Random Energy
Model (REM) \cite{Derrida81} is the simplest archetype.

The transition to the soft glass phase is a continuous one, corresponding to
full replica symmetry breaking (FRSB). This is more in line with recent
scenarios proposing a freezing that proceeds gradually from small scales to
larger and larger structures \cite{Copolymerdyn,CopolymerIsing}. In a series of papers exploiting a Gaussian
variational technique to deal with the dynamics of heteropolymers, copolymers
in particular, a much richer phase diagram was proposed, where the ultimate
REM-like folding to a unique ground state is preceded by a less structured but
still frustrated glassy phase \cite{Copolymerkinetics,CopolymerPhasedia,Timoshenko98}. As for the glass transition, the random
copolymer was proposed to be in the same universality class as the Ising spin
glass \cite{CopolymerIsing}, which would imply a continuous transition with a
full breaking of the replica symmetry.
 
Beside providing an alternative and well controlled analytical 
approach, our cavity analysis adds 
to the above pictures in that it highlights the dependence of the scenario 
to be expected on the correlations of the monomer sequences.

In order to keep the computations more transparent we avoid 
here the use of replicas (although it would be possible to write all of 
the ensemble-averaged
cavity equations using replicas), but we keep to the traditional
replica vocabulary of 1RSB and FRSB to denote the two types of transitions.

We will apply here the general method to treat Markov-correlated
sequences. However, a much wider range of possible applications of
this technique is open. 

The paper is organized as follows: In Section \ref{cavityapproach} 
we define the lattice model and review the treatment of polymers 
in the grand canonical ensemble. We then 
introduce the basic ideas of the Bethe approximation and
discuss the $\Theta$-collapse from the random coil to 
the liquid globule phase. Section \ref{glassphase} 
discusses the shortcomings of the liquid solution and
generalizes the method to the case where many pure states exist (as 
typically in a glassy phase). 
In particular, we propose a set of local order parameters that
allow to distinguish both theoretically and experimentally between two 
different types of glass transitions. In Section \ref{MethodsSection}
we describe some basic tools for analyzing the glass transition.
We present a local stability criterion for the liquid phase and
the 1RSB cavity equations which are used to describe the glassy phase. 
This formalism is illustrated 
in  Section \ref{alternatingsequences} by considering the exemplary cases of  alternating sequences with attractive or repulsive interactions of like monomers.

It turns out that the two types of
interactions imply very different phase transitions: either a continuously 
emerging ``soft'' glass phase or the ``standard'' discontinuous 
freezing transition.
These two scenarios are found in the study of 
Markovian chains in Sec. \ref{MarkovianSection}.
The properties of the strongly frozen phase is analyzed in 
Section \ref{frozenphase} by focusing on maximally compact conformations.
We conclude with a summary of 
our results and a discussion of their relevance for protein 
folding. 
Several technical developments are included in the seven appendices.
%
%

\section{The cavity approach to heteropolymers}
\label{cavityapproach}

In this Section we  describe the type of heteropolymer models which we 
shall study.
We derive their phase diagram under the assumption that 
the polymer is ``liquid'' meaning 
that any 
statistically relevant conformation is dynamically accessible to the molecule.
In replica jargon this corresponds to assuming replica symmetry.
The next sections will render more precise the regions of the phase 
diagram where this liquid phase is stable and corresponds to the 
physically relevant state.
%
%
\subsection{The lattice polymer model}
\label{DefinitionSection}

Our starting point is the standard model of lattice polymers 
\cite{Go75,ChanDill89},
which we generalize for polymers living on a general graph ${\cal G}$.
We  denote by $i,j,\dots\in {\cal V}$ the vertices of ${\cal G}$ (with
$|{\cal V}|=V$), and by $(i,j),\dots\in {\cal E}$ the edges of ${\cal G}$.
Let $\uo = (\omega_1\dots\omega_N)$, $\omega_a\in {\cal V}$ denote a 
self-avoiding walk (SAW) of length $N$ on  ${\cal G}$. 
The position of a monomer along the chain is denoted by 
$a,b,\dots\in \{1\dots N\}$, and we assume an interaction matrix
$e_{ab}$ to be assigned. The corresponding energy reads:
\begin{eqnarray}
H_N(\uo) = \sum_{(a,b)|(\omega_a,\omega_b)\in {\cal E}}\,  e_{ab} ,
\label{Hamiltonian}
\end{eqnarray}
where the sum runs over couples of non-consecutive monomers which are nearest neighbors on the lattice. 

The choice of the matrix $e_{ab}$ is crucial. The standard homopolymer model
is recovered by setting $e_{ab} = e_0$. A popular model in heteropolymer
studies is the {\it random bond} model {\cite{ShakhnovichGutin89c} which
assumes the $e_{ab}$ to be independent identically distributed (i.i.d)
quenched random variables. In this work we study the more realistic case where the
interaction energies are determined by the underlying monomer sequence. The
sequence will be given by $\{\sigma_1,\dots,\sigma_N\}$, with $\sigma_a\in
{\cal A}$ being the type of the monomer at position $a$ in the sequence. The interaction energy of two
monomers is assumed to depend only upon the monomer type: $e_{ab} =
E_{\sigma_a\sigma_b}$. In particular, we shall focus on copolymers (although
the approach is general) where there are only two types of monomers: ${\cal A}
= \{ A,B\}$. Interaction matrices $E_{\sigma,\sigma'}$ of particular interest are:
\begin{itemize}
\item \underline{The HP model.} $A$ and $B$ monomers represent (respectively)
hydrophobic and polar aminoacids, and the interaction matrix is chosen 
accordingly, e.g., $E_{AA}=-1$, $E_{AB} = E_{BB}= 0$. This is a popular toy model for protein folding \cite{HPmodel}.
\item \underline{The polyampholyte.} $A$ and $B$ are supposed to carry 
screened charges which suggests $E_{AA}=E_{BB}=+1$ and $E_{AB} = E_{BA} = -1$. 
Sometime we shall refer to this interaction matrix as the antiferromagnetic 
(AF) model. 
\item \underline{The symmetrized HP model.} We take 
$E_{AA}=E_{BB}=-1$ and $E_{AB} = E_{BA} = +1$. This is the standard model for copolymers with monomers that have a tendency to segregate \cite{SfatosShakhnovich97}. We shall refer to it as the ferromagnetic (F) model.
\end{itemize}

As for the graph ${\cal G}$ we shall consider two particular cases:
$(i)$ A $V$-sites portion of the $d$-dimensional cubic lattice. $(ii)$ 
A $V$-sites Bethe lattice, i.e., a random lattice with connectivity
$(k+1)$.
Its interest stems from the observation that, in the thermodynamic limit, 
our mean-field calculations are exact on such a graph.

Both for our analytical computations and for the simulations on the Bethe 
lattice we shall need to consider periodic sequences with period $L$: 
$\sigma_{i} = \sigma_{i+L}$. 
The complete sequence is therefore determined by its first period
$(\sigma_1\dots \sigma_L)$. Hereafter, we shall use 
the shorthand notation ``monomer $a$'' to refer to 
all monomers in positions $a+nL$ with integer $n$. Furthermore, 
monomer indices always should be read modulo $L$.
We expect the non-periodic case to be recovered in the
$L\to\infty$ limit, even if this limit is taken
 {\it after} the limit $N,V\to\infty$.

The random-bond model is obtained in the $|{\cal A}|=L\to\infty$
limit by setting $\sigma_1\ne\sigma_2\ne\dots\ne\sigma_L$, and taking the
$E_{\sigma,\sigma'}$ to be i.i.d. random variables.

In order to understand the influence of the correlations in the 
sequence of monomers, we shall consider Markovian random 
copolymer chains in the large $L$ limit. 
In these chains the probability of a monomer to be of a certain type 
depends only on the preceding monomer in the sequence. 
For the sake of simplicity we assume the two types of monomers to
occur with the same frequencies.
The statistical ensemble of the chains is then fully characterized by the 
probability $\pi \in [0,1]$ of a monomer to be of the same type as the 
preceding one.

We study the system at thermal equilibrium at a temperature $T=1/\beta$.
We  define a canonical free energy density as
\begin{eqnarray}
-\beta\, f_{L}(\beta,\rho) = \lim_{\stackrel{N,V\to\infty}{N=\rho V}}
\frac{1}{V}{\mathbb E}_{\cal G}\log \left(\sum_{\uo}\ex^{-\beta H_{N}(\uo)}
\right)\, ,
\end{eqnarray}
and its grand-canonical counterpart
\begin{eqnarray}
-\beta\, \omega_{L}(\beta,\mu) = \lim_{V\to\infty}
\frac{1}{V}{\mathbb E}_{\cal G}\log \left(\sum_{N\ge0} \ex^{\beta\mu N}
\sum_{\uo}\ex^{-\beta H_{N}(\uo)}\right)\, ,
\end{eqnarray}
where the expectation value ${\mathbb E}_{\cal G}$ is taken with respect to 
the graph ensemble (whenever ${\cal G}$ is a random graph). 
The $L\to\infty$ limit, and the expectation with 
respect to the sequence $(\sigma_1\dots \sigma_L)$ are (eventually)
taken afterwards.

The two free energies defined above satisfy the usual Legendre transform 
relation $\omega_L(\beta,\mu) = f_L(\beta,\rho)-\mu\rho$.
In order to describe free polymers (in equilibrium with the solvent) the 
chemical potential has to be adjusted to the critical value $\mu_c$  
such that $\omega_L(\mu_c)=0$ \cite{DeGennes75}. In the 
grand-canonical picture this critical line corresponds to a phase transition 
between an infinitely diluted phase for $\mu<\mu_c$ and a dense 
phase with non-vanishing osmotic pressure for $\mu>\mu_c$. 
If this phase transition is continuous, 
the density on the coexistence line vanishes, 
while it is finite if the transition is first order. On this coexistence line,
the tricritical point where the nature of the transition 
changes is nothing but the $\Theta$-point where the collapse of the unconstrained 
polymer takes place.

In a homopolymer, the above description captures the essential of 
the phase diagram \cite{Bethehomopolymer}. However, in a heteropolymer, the low temperature dense 
phase will be strongly influenced by the sequence heterogeneity. Due to 
the connectivity of the polymer chain it is in general impossible to 
find a compact folding where all interactions are favorable. The system is frustrated, and a glass transition will take place at 
sufficiently low temperature. 
%
%

\subsection{The Bethe Peierls approximation}

As already mentioned, the Bethe approximation
is asymptotically exact on locally tree-like graphs.
Following \cite{cavity0}, we define a Bethe lattice as a
random lattice with fixed connectivity. Such a lattice is locally tree-like
since the typical loop size diverges as $\log V$ with the lattice size.  
In order to handle the heteropolymer problem on a $d$-dimensional
hypercubic lattice within the Bethe approximation, our 
 approach idealizes 
the graph as a Bethe lattice with the same connectivity, $k+1 = 2d$.
 
The local tree structure of the graph can be exploited in
a recursion procedure. Suppose for a moment that the lattice is a tree, and 
let us single out a single branch of the tree which is rooted at one 
`cavity site' $0$ having only $k$ neighbors $i=1,..,k$.
In the absence of $0$, the branch would become a collection
of $k$ other branches, rooted at $i=1,..,k$. This structure allows for a recursive
computation of the probabilities of the polymer's conformations on the tree.

\begin{figure}
	\resizebox{9 cm}{!}{
  \includegraphics{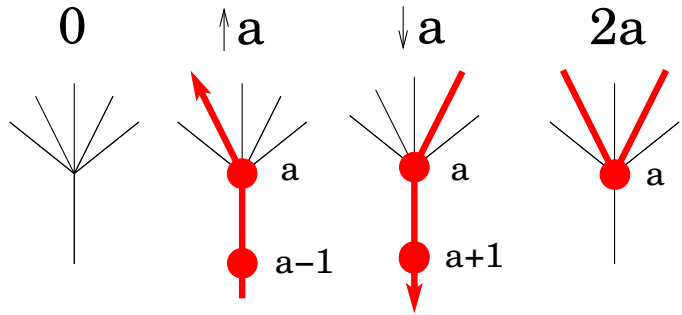}}
\caption{Possible conformations of a site (or oriented edge) on the 
regular Bethe lattice. 
The cavity site is considered as the root of a branch with $k$ leaves (here $k=5$). 
The thick lines and filled circles represent the
chain backbone and monomers.}
\label{figconformations}
\end{figure}

  We first list  the  possible local conformations of the cavity site $0$ 
 in its branch (see Fig.~\ref{figconformations}). 
$(0)$: the site is unoccupied; $(\ua a)$ or $(\da a)$: the site is 
occupied by the monomer $a$ and the backbone continues towards the 
remainder of the tree, with monomer $a-1$ or $a+1$, respectively; 
$(2a)$: the site is occupied by monomer $a$, but the polymer returns 
back to the leafs. 
(On a real tree the 
parts of the polymer on different branches  are necessarily 
disconnected. However, on the Bethe lattice this is no longer the case
and the polymer may be present on more than two leaves.)

For each  local conformation $ \alpha \in \{0, \ua a, \da a, 2a\}$
of the root site $0$, we denote by $p_{\alpha}^{(0)}$ the corresponding
probability (as given by the Boltzmann measure). The ($3L+1$) dimensional vector of weights 
$ {\bf p}^{(0)}$,
with components $p_{\alpha}^{(0)}$, can be expressed in terms of the 
corresponding $k$ 
weight vectors ${\bf p}^{(i)}$ on the neighboring sites. Note that
$ p_{\alpha}^{(i)}$ is the Boltzmann weight for the configuration 
$\alpha$ on $i$ when the site $0$ is absent. We will refer to these weight 
vectors on root-sites as cavity  fields. 

The  mapping between cavity fields, 
$ {\bf p}^{(0)}=I\left[{\bf p}^{(1)},...,{\bf p}^{(k)}\right]$, 
can be written explicitly as:
\begin{eqnarray}
\label{cavity0}
p_0^{(0)} &=&C^{-1}\prod_{i=1}^{k} \psi_0^{(i)},\\
\label{cavityup}	
p_{\ua a}^{(0)} &=&C^{-1}e^{\beta \mu}\sum_{i=1}^{k} p_{\ua a+ 1}^{(i)} 
\prod_{j\ne i} \psi_a^{(j)},\\
\label{cavitydown}	
p_{\da a}^{(0)} &=&C^{-1}e^{\beta \mu}\sum_{i=1}^{k} p_{\da a- 1}^{(i)} 
\prod_{j\ne i} \psi_a^{(j)},\\
\label{cavity2}
p_{2a}^{(0)} &=& C^{-1}e^{\beta \mu}\sum_{i_1\ne i_2} p_{\da a-1}^{(i_1)} 
p_{\ua a+1}^{(i_2)} \prod_{j\ne i_1,i_2} \psi_a^{(j)},
\end{eqnarray}
where $C\equiv C[\{{\bf p}^{(i)}\}]$ is a normalization constant which 
enforces the condition $\sum_{\alpha}p^{(0)}_{\alpha} =1$ 
and we have introduced the quantities
\begin{eqnarray}
\psi_0^{(i)}=p_0^{(i)}+\sum_{a'=1}^L p_{2a'}^{(i)}\, ,\;\;\;\;\;
\psi_a^{(j)}=p_0^{(j)}+\sum_{a'=1}^L p_{2a'}^{(j)}e^{-\beta e_{aa'}}.
\end{eqnarray}

The full lattice is built by merging $k+1$ branches. Therefore,
once the cavity fields have been computed, one can express 
any local quantity using the neighboring cavity fields. 
The monomer density $\rho^{(i)}$ at site $i$ is a function  
of the $k+1$ cavity fields ${\bf p}^{(j)}$
on the $j=1,...,k+1$ neighboring sites of $i$ (recall that ${\bf p}^{(j)}$
gives the probability of a local conformation on $j$ in the absence of $i$):
\begin{eqnarray}
\rho^{(i)}= \sum_{a=1}^{L}\sum_{j_1\neq j_2}\frac{ p^{(j_1)}_{\ua a+1} 
p^{(j_2)}_{\da a-1} \prod_{j\ne j_1,j_2} \psi^{(j)}_a}
{w_s^{(i)}({\bf p}^{(1)},...,{\bf p}^{(k+1)})}\, ,\label{Density}
\end{eqnarray}
where we have defined the normalization constant
\begin{eqnarray}
w_s^{(i)}({\bf p}^{(1)},...,{\bf p}^{(k+1)})&=& \prod_{j=1}^{k+1}\psi_0^{(j)}
+e^{\beta \mu}\sum_{a=1}^{L} \sum_{j_1\ne j_2} p_{\da a-1}^{(j_1)} 
p_{\ua a+1}^{(j_2)} \prod_{j\ne j_1,j_2} \psi_a^{(j)}\, .
\label{SitePartitionFunction}
\end{eqnarray}

The internal energy $u_{ij}$ of a link $(i,j)$ can be written in terms of the
cavity fields on $i$  and $j$ (giving the probabilities
of local conformations on $i$ and $j$ in the absence of  the link $(i,j)$):
\begin{eqnarray}
u_{ij} = \sum_{a,b=1}^L e_{ab}\, n_{ij}(a,b)\, ,
\;\;\;\;\;\;\; n_{ij}(a,b) = \frac{p^{(i)}_{2a}p^{(j)}_{2b}
\, \ex^{-\beta e_{ab}}}{w_l^{(ij)}({\bf p}^{(i)},{\bf p}^{(j)})}\, ,\label{Energy}
\end{eqnarray}
where $n_{ij}(a,b)$ is the probability of having a contact between two
monomers $a$ and $b$ along the link $(ij)$ of the graph. The normalization $w_l({\bf p}^{(i)},{\bf p}^{(j)})$ 
is given by
\begin{eqnarray}
\hspace{-1cm}w_l^{(ij)}({\bf p}^{(i)},{\bf p}^{(j)})
&=& p_0^{(i)}p_0^{(j)}+\sum_{a,b=1}^{L}p_{2a}^{(i)}
p_{2b}^{(j)}\ex^{-\beta e_{ab}}
+\sum_{a=1}^L \left(p_0^{(i)}p_{2a}^{(j)}+p_{2a}^{(i)}p_{0}^{(j)}
+ p_{\da a-1}^{(i)} p_{\ua a}^{(j)}+p_{\ua a}^{(i)} p_{\da a-1}^{(j)}
\right)\, .
\label{LinkPartitionFunction}
\end{eqnarray}

For each edge $(i,j)$ of a given graph, one can introduce 
a pair of cavity fields, describing respectively the probability of local configurations of the two points $i$ and $j$ in the absence of the edge $(i,j)$.
One can write a Bethe free energy, which is a functional of 
all these cavity fields and has Eqs. (\ref{cavity0})-(\ref{cavity2}) as 
stationarity conditions. It reads
\begin{eqnarray}
V\beta\omega[\{{\bf p}^{(i)}\}] = -\sum_{i\in{\cal V}} \log[ w_s^{(i)} ] +
\sum_{(ij)\in{\cal E}}\log[ w_l^{(ij)}]\, ,\label{FreeEnergy}
\end{eqnarray}
where  $w_s^{(i)}$ and $w_l^{(ij)}$ are the expressions given in
 (\ref{SitePartitionFunction}) and 
(\ref{LinkPartitionFunction}), respectively.
Notice moreover that the density (\ref{Density}) and the internal energy
(\ref{Energy}) can be obtained by differentiating the Bethe free energy with 
respect to the chemical potential $\mu$ and the inverse
temperature $\beta$.

It is easy to show that the above expressions are exact if the 
graph ${\cal G}$ is a tree.
On a general lattice it holds approximately to the extent that one 
can neglect the correlations between the fields on the $k+1$ neighbors
of any site $i$, once the site $i$ itself has been deleted. 

On a Bethe lattice, since the typical loop size diverges as $\log V$ 
in the large-$V$ limit, these $k+1$  sites neighbors of $i$ are generically distant
from each other, when $i$ is absent. Therefore the correlations
of their fields can be beglected, if the system is in
a single pure state:
at low temperature the Gibbs measure usually has to be decomposed 
into pure states, within which the  correlations between two sites
 decay with their distance along the graph. 
We thus expect the above cavity approximation to become asymptotically exact,
insofar as cavity fields are computed within one pure state.
%
%
\subsection{The liquid solution and the $\Theta$-point}

Both on the random Bethe lattice and on the $d$-dimensional cubic graph,
each site has generically the same environment within any distance $R$ 
(as long as $R$ is kept finite in the $V\to\infty$ limit).
A liquid phase is therefore expected to enjoy translational invariance
and will be described by a set of fields $p^{(i)}_\alpha$ that 
is independent of the site. We thus look for a fixed point 
$p_\alpha^{(i)}\equiv p_\alpha^{*}$ of the recursions 
(\ref{cavity0})-(\ref{cavity2}).

It turns out that the liquid solutions can be found by 
solving a system of $|{\cal A}|+2$ non-linear equations,
$|{\cal A}|$ being the number of monomer species in the model. 
This is a great complexity reduction with respect to the $3L+1$
equations (\ref{cavity0})-(\ref{cavity2}).
The task can be further simplified by using particular symmetries
of the interaction matrix. This is, for instance, the case of
the F- and AF-models defined in Sec. \ref{DefinitionSection},
which are symmetric under the interchange $A\leftrightarrow B$.
We refer to App. \ref{LiquidApp} for a detailed discussion  of
how the solution is obtained.

As shown in Appendix \ref{LiquidApp} all the  thermodynamic quantities
depend upon the sequence $(\sigma_1\dots\sigma_L)$ only
through the fractions $\nu_{\sigma}$ of monomers of type $\sigma$.
As a byproduct, the $L\to\infty$ limit can immediately be taken.
The physical meaning of this result is easily understood.
In the liquid phase, the correlations induced by the sequence
play some role just along the chain, and their net effect
vanishes at large distance.
In particular, the monomer  $a$ is surrounded by 
a certain fraction of monomers of type $\sigma'$ which only depends 
on the type of $a$, $\sigma_a$ (apart from the sites occupied by the 
monomers $a-1$ and $a+1$, of course).

Let us now discuss the various solutions of liquid type.

The random coil phase is described by the trivial solution 
$p^*_{\alpha}=\delta_{\alpha,0}$, which exists for any choice of the
parameters.
This phase has vanishing grand potential $\omega$ and density $\rho$. At high
temperatures this is the only solution when $\mu$ 
is smaller than the critical
chemical potential $\mu_c$ given by $\exp(\beta\mu_c)=1/k$.
At $\mu_c$   a non-trivial solution
emerges continuously. The latter describes a liquid phase under
pressure ($\omega>0$ for $\mu>\mu_c$) with a density that vanishes 
on approaching the critical line.

The collapse of a free polymer from the random coil state to the
liquid globule occurs at the so-called $\Theta$-point. In the
grand-canonical description, it appears as the tricritical point on
the line $\exp(\beta\mu)=1/k$. 
Expanding around $p^*_{\alpha}=\delta_{\alpha,0}$, one obtains
the following relation which determines the $\Theta$-point temperature
\begin{eqnarray}
\label{thetapoint}
\sum_{\sigma,\tau\in{\cal A}} \nu_\sigma \nu_\tau\, 
e^{-\beta_{\Theta}E_{\sigma\tau}}
=\frac{k}{k-1}\, ,
\end{eqnarray}
see App. \ref{LiquidApp}.
This result has previously been obtained within the framework 
of the standard cluster variational method \cite{Pretti02}.
At temperatures below the $\Theta$-point, 
 $\beta>\beta_{\Theta}$, the grand-canonical phase transition
becomes first order (see Fig.~\ref{liquidPhasedia}).
\begin{figure}
	\resizebox{9 cm}{!}{
  \includegraphics{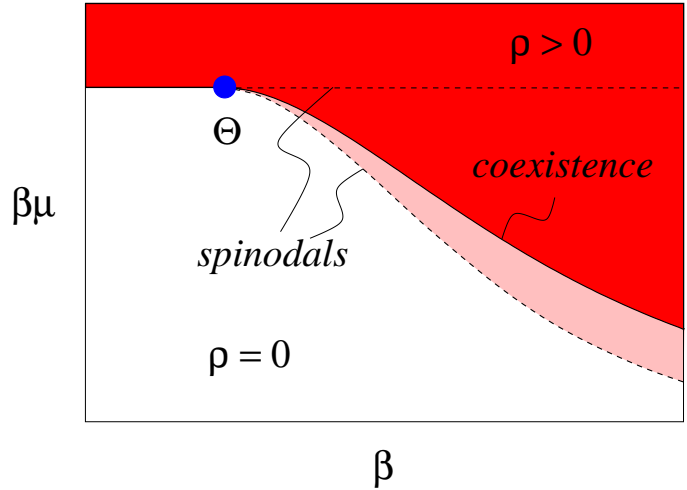}}
\caption{The phase diagram corresponding to the liquid (translation invariant)
 solution in the grand canonical ensemble. Above the $\Theta$-temperature, 
$\beta<\beta_\Theta$, the phase transition from the random coil phase 
($\mu<\mu_c$) to the globule solution with finite density ($\mu>\mu_c$) 
is continuous. At low temperatures,$\beta>\beta_\Theta$ , the transition becomes first order 
and is accompanied by two spinodals. The globule solution on the critical 
line describes a free polymer in coexistence with the surrounding pure solvent. The free polymer undergoes a collapse transition at the $\Theta$-point}
\label{liquidPhasedia}
\end{figure}
The critical line $\mu_c(\beta)$ is obtained by equating the grand potentials 
in the coil and globule phases, i.e., by solving $\omega=0$ for the 
globule solution. The  density, internal energy, and free energy 
are obtained by plugging
the globule solution $p^*_{\alpha}$ into Eqs. (\ref{Density}), (\ref{Energy}),
(\ref{FreeEnergy}).

In the low temperature region $\beta>\beta_{\Theta}$, the dense solution 
can be continued to values of the chemical
potentials smaller than the critical one $\mu_{\rm c}(\beta)$,
and ceases to exist on a spinodal line. 

Likewise, the trivial dilute solution stays locally stable 
beyond the coexistence line up to the spinodal $\exp(\beta\mu)=1/k$. 

The above results compare reasonably with the outcomes of 
numerical simulations on $d$-dimensional lattice. For instance, the
homopolymer $\Theta$-point on the cubic lattice given by $T_{\Theta} = 1.50$
for $d=2$ \cite{GrassbergerHegger95b}, $3.716(7)$ for $d=3$ 
\cite{TesiJanse96}, and $5.98(6)$ ($d=4$) \cite{PrellbergOwczarek00}. 
Moreover the authors of Ref. \cite{diamond} found $T_{\Theta} = 2.25(10)$ on the
three-dimensional diamond lattice (connectivity $k+1=4$).
These results should be compared with the outcome of the Bethe approximation,
cf. (\ref{thetapoint}), which yields 
$T_{\Theta,{\rm Bethe}} \approx 2.4663035$
(for $k=3$), $3.4760595$ ($k=4$), $4.4814201$ ($k=5$). 
  $6.4871592$ ($k=7$).
As for heteropolymers, the authors of 
Ref.~\cite{KantorKardar94,GrassbergerHegger95}
estimated  $T_{\Theta} \approx 1.2$ both for the F- and AF-models
of Sec.~\ref{DefinitionSection} in $d=3$. This result is compatible with
$T_{\Theta}=1/\log(2)\approx 1.442695$ which comes out of 
Eq. (\ref{thetapoint}).

Finally, several numerical studies \cite{Monari99,BastollaGrassberger01}
have focused on the $\Theta$-point
of random bonds models, and have argued that its location is extremely 
well approximated by an annealed computation. Once again, this 
confirms that Eq.~(\ref{thetapoint}) is a reasonable approximation
(the random-bond model is recovered by setting $|{\cal A}|=L$, 
$\nu_{\sigma}=1/L$ and $E_{\sigma\tau}$ i.i.d.'s random variables). 
This is also related to the numerical finding that the global 
collapse in protein folding dynamics is essentially unsensitive 
to the specific structure of the sequence, but only depends on its 
global composition \cite{BryngelsonReview95}.
%
%
\section{Glass phases}
\label{glassphase}
If we follow the entropy density $s(\beta)$ of the liquid solution as a
function of temperature, we find that in any heterogeneous sequence $s(\beta)$
turns negative at sufficiently low temperatures. This indicates the existence
of a phase transition to a glass phase which breaks the translational
invariance. 

As we will show, this glass transition can be of two types.
In certain sequences the ``entropy crisis'' is
preceded by a local instability of the cavity recursions
(\ref{cavity0})-(\ref{cavity2}) around the liquid fixed point $p^*_{\alpha}$.
This implies the divergence of a properly defined spin-glass susceptibility
and signals a continuous glass transition
towards a phase with fully broken replica symmetry.

In other sequences, and in the Gaussian random bond model, this 
local instability is irrelevant since it occurs - if at all - in 
the region of negative entropy of the liquid globule. The glass transition 
is thus necessarily discontinuous (1RSB), as was predicted 
from replica calculations 
for the random  bond model \cite{ShakhnovichGutin89a}.

Dealing with the glass phases requires some modifications of the
simple Bethe Peierls approximation which we have been using so far.
In this section we will describe first some general properties
of the glass phases, and explain the general technical tools that can be used
to study glass transitions using the cavity method.

\begin{figure}

\epsfig{figure=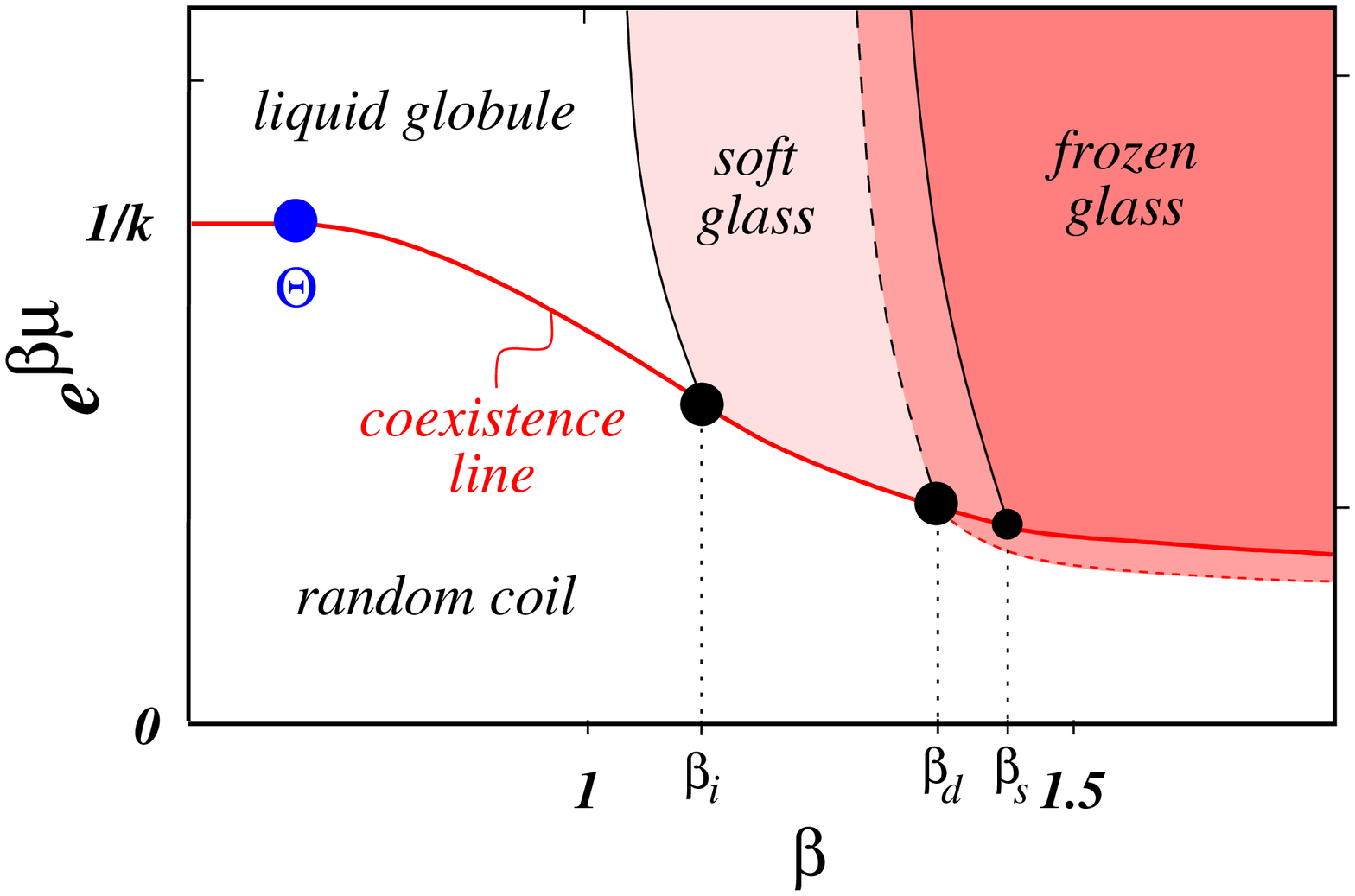,width=0.6\linewidth} 
\epsfig{figure=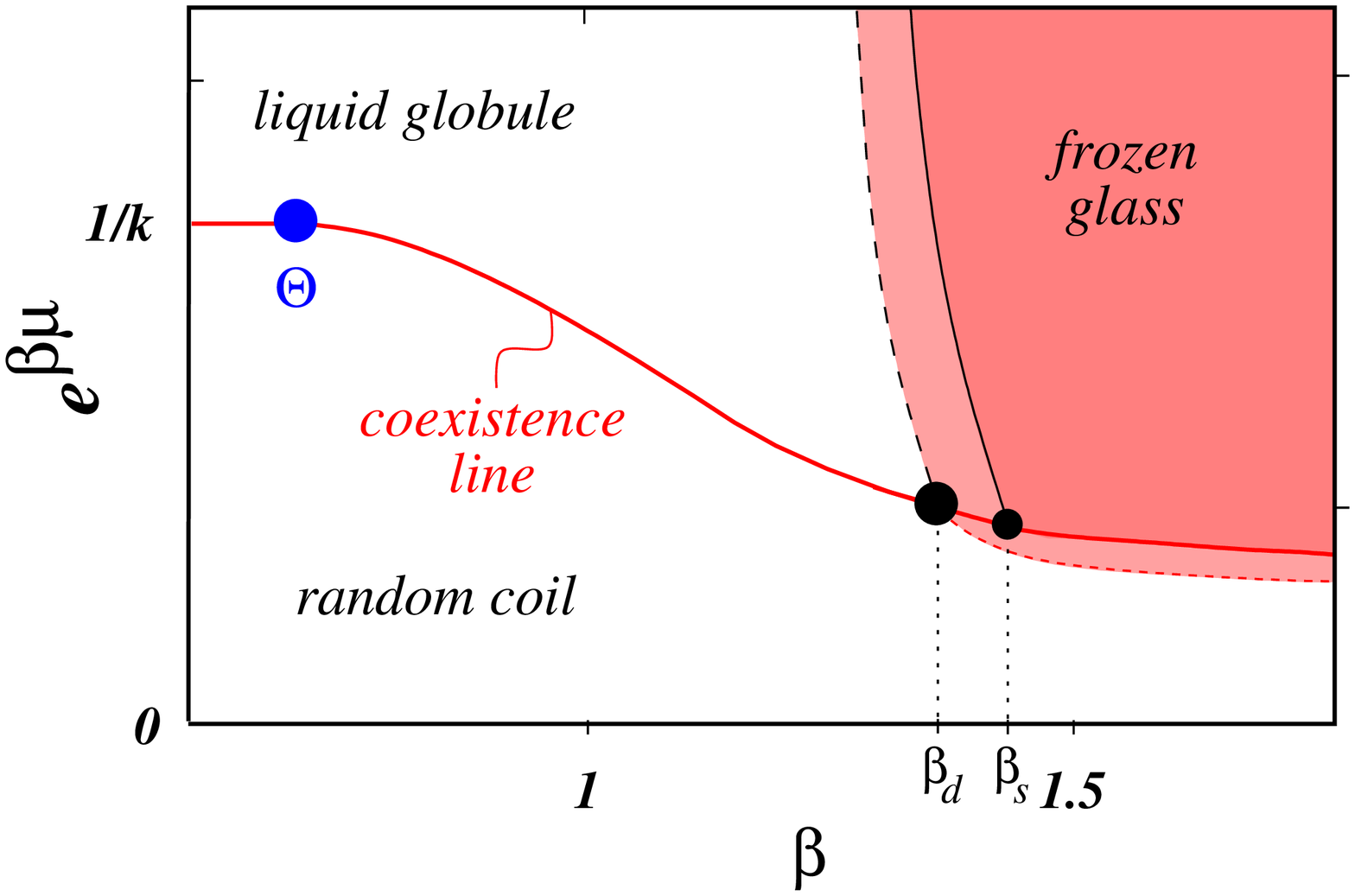,width=0.6\linewidth}

\caption{Schematic phase diagram of copolymers as a function of 
inverse temperature $\beta$ and chemical potential $\mu$. 
A polymer in equilibrium with the solvent is described by the coexistence 
line. Beyond the $\Theta$-point, $\beta>\beta_\Theta$ it is in a collapsed 
phase with a finite density.
Depending on the sequence correlations of the copolymer there may be 
a local instability of the liquid (dash-dotted line), giving rise to a continuous glass 
transition at $\beta_i$ (see upper graph). 
In the absence of a local instability down to a critical temperature in 
the range of  $\beta\approx 1.23$, a discontinuous glass transition 
will take place. 
The thermodynamic (static) phase transition at $\beta_s$ is 
preceded by a dynamic glass transition at $\beta_d$ where the phase 
space splits up into different pure states. In the glass phase, the 
critical chemical potential depends on whether the dynamically relevant 
threshold states (dashed line) or the states dominating the static 
equilibrium (solid line) are described.  
}
\label{phasedia}
\end{figure}

%
%
\subsection{Proliferation of pure states}
\label{sect:mdef}
In a glassy phase, the space of conformations is expected to split up in a 
multitude of pure states that are separated by large free energy barriers.
The slowest time scale of the system, corresponding to jumps between
pure states, increases dramatically. 

In mean field approximation, or on the Bethe lattice, 
this time scale diverges and ergodicity is
broken at the ``dynamic'' phase transition. The system eventually
undergoes a ``static'' phase transition (with a non-analyticity
in the thermodynamic potentials) at a lower temperature \cite{KirWol87,BouchaudCugliandolo98}.

In a finite-dimensional model the ``dynamic'' phase transition
becomes a crossover where the nature of the most important dynamical
processes changes. Whether  the ``static''
phase transition  survives 
in a given model, or not, is not known in general.
We shall not enter this
dispute here since we have little to say about it. In any case,
the mean-field-like Bethe approximation, assuming the existence of many pure states, yields some useful insight on the glass phase.

Within one pure state, the conformational probabilities on a given site are
well-defined \cite{cavityT,cavity0}. However, there is no reason to assume the
equality of local fields on different sites. Rather one expects that in a
given pure state the sites will have different preferences for certain polymer
conformations.

To proceed, one has to use a statistical description of local fields.
We shall not explain here all the details of this description, but 
just give the main definitions and refer the reader to \cite{cavityT,cavity0} 
for detailed discussions.
In a glassy phase, the number of pure states $\mathcal{N}_V(\omega)$ 
increases exponentially with the volume of the system. 
The complexity $\Sigma(\omega)$ is  the 
monotonously increasing, concave function defined 
by $\mathcal{N}_V(\omega)\sim\exp(V\Sigma(\omega))$.
The natural order parameter is the distribution of local fields over 
the pure states $\gamma$ whose free energy 
density $\omega_\gamma$ is fixed to a value $\omega_0$: 
\begin{equation}
\label{rho(p)}
\rho({\bf p}) \propto \sum_{\gamma}\delta({\bf p}-{\bf p}^{(i,\gamma)}) 
\delta(\omega_\gamma-\omega_0).
\ee
An alternative description consists in using a Legendre transformation of the
complexity, by introducing the parameter $m=(1/\beta) \Sigma'(\omega_0)$
and working at fixed $m$ instead of fixed $\omega_0$ 
\cite{Monasson95}. This computation is 
equivalent to a 1RSB scheme with 
Parisi parameter $m$. From the free energy at fixed $m$, $\phi_1(m)$,
 the complexity $\Sigma(\omega)$
is obtained through the Legendre transform: 
$m\beta\phi_1(m)=m\beta \omega-\Sigma(\omega)$.

In a system with a discontinuous (1RSB) glass transition, this approach gives
a full description. The complexity is strictly positive in the interval
$\omega_{\rm s}<\omega <\omega_{\rm d}$, corresponding to the interval $m_{\rm
d}<m <m_{\rm s}$ in the 1RSB parameter. The thermodynamically dominant
metastable states are obtained by minimizing the one-replica free energy
$\omega-\beta^{-1}\Sigma(\omega)$. In an intermediate temperature regime
$T_{\rm s}<T<T_{\rm d}$, the minimum is attained for some free energy
$\omega_*$ (corresponding to $m_*=1$), with $\omega_{\rm
s}<\omega_*<\omega_{\rm d}$. Below the glass transition, $T<T_{\rm s}$, the
minimum is attained at the lower edge $\omega_*=\omega_{\rm s}$ 
(with $\Sigma(\omega_*)=0$), corresponding to the 1RSB parameter $0<m_*<1$.

In a system with a continuous glass transition (FRSB), the full solution
should involve grouping states into clusters, and clusters into
superclusters, building up a continuous ultrametric hierarchy. The approach
above amounts to a 1RSB approximation of this full structure, and we shall not
attempt 
to go beyond this level of approximation.

\subsection{Order parameters}
\label{Freezing}
In this section we present two types of order parameters which can be used to
identify the glass phase.

For a polymer in Euclidean space, described by the position $\vec R_i$ of
monomer $i$, let us consider two replicas of the polymer in the same 
pure state. In the glass phase, provided the global rotation 
symmetry is broken, the local conformation of the two polymers will have 
a certain tendency to be the same while the liquid phase is completely 
disordered in this respect. In order to measure
this effect, we introduce the scalar product of the distance vectors between
nearby monomers in the replicas (1) and (2): 
\be
F_d^{(1,2)}=\sum_i(\vec{R}^{(1)}_{i+d}-\vec{R}^{(1)}_{i})\cdot
(\vec{R}^{(2)}_{i+d}-\vec{R}^{(2)}_{i}) \ . 
\ee 
We shall be interested in
computing the average of this quantity when the replicas are constrained to remain
in the same pure state. More precisely, we want to evaluate 
\be
\<F_d^{(1,2)}\>_{\rm state} = \sum_{\gamma}w_{\gamma}\<F_d^{(1,2)}\>_{1,2 \in
\gamma} \,,
\label{orpar_space}
\ee 
where we average over all states $\gamma$ with their 
Boltzmann weigth $w_\gamma$. This quantity is
accessible numerically. We consider a polymer which is thermalized at time $0$
in a configuration $\vec R_i(t=0)$. We let it evolve for a time $t$, to a
configuration $\vec R_i(t)$. The order parameter is given by the quantity
\be
\<F_d^{(1,2)}\>_{\rm state} = \<\frac{1}{t_{\rm MAX}}
\int_0^{t_{\rm MAX}}dt\,\frac{1}{N}\sum_i (\vec{R}_{i+d}(t)-\vec{R}_{i}(t))
\cdot (\vec{R}_{i+d}(0)-\vec{R}_{i}(0))\>_{\{R_i(t=0)\}}\,,
\ee
evaluated over timescales $t_{\rm MAX}$ which are
large but much smaller than the typical timescale for interstate transitions or even full equilibration (in
particular much smaller than the time scale for diffusion or rotation of the
polymer, which diverges with $N$).

A simpler order parameter can be defined by first introducing, 
on each site $i$ of the  lattice, the quantity $s_i$ which takes 
the value $s_i=1$ if the site is occupied by a monomer $A$, $s_i=-1$ 
if there is a $B$ monomer and $s_i=0$ if the site is empty.
Then the overlap between two configurations 1 and 2 of 
the polymer can be defined as
\begin{eqnarray}
q^{(1,2)}_{AB} = \frac{1}{V}\sum_{i=1}^V s^{(1)}_is^{(2)}_i\, .
\label{ABOverlapDefinition}
\end{eqnarray}
Again, one can compute the typical distance 
$\<q^{(1,2)}_{AB}\>_{\rm state}$ between 
two conformations in the same state by recurring to dynamical 
simulations.

Notice that both $q^{(1,2)}_{AB}$ and $F_d^{(1,2)}$ define a notion of
distance (or similarity) between polymer configurations. 
However, they describe two complementary aspects of the polymer: 
$q^{(1,2)}_{AB}$ essentially characterizes the bias of single sites 
towards a specific monomer type, whereas the order parameters $F_d^{(1,2)}$ 
measure the \textit{conformational} similarity of the replicas in the 
vicinity of a given site, once the monomer on that site has been fixed. 
They measure the freezing of the local degrees of freedom of the polymer's
 backbone, similarly to the approach of 
\cite{Copolymerkinetics,CopolymerPhasedia,Timoshenko98}. In contrast 
the parameter $q^{(1,2)}_{AB}$ is hardly sensitive to the geometric 
constraints induced by the backbone.

A dynamical evaluation of the above order parameters is particularly 
convenient on finite-dimensional lattices. Notice that the equilibrium
probability for two independent replicas to have a finite overlap 
$q^{(1,2)}_{AB}$, vanishes with the volume of the lattice because of 
translation invariance.

On the Bethe lattice it is more natural to work at a finite
monomer density, (see  Sec. \ref{NumericalBethe}). 
In this case, the random structure of the lattice acts as a 
``pinning field'', and two replicas of the same system typically have 
a finite overlap.
Following the practice from spin-glass theory,
we shall measure the probability distribution of the quantity 
(\ref{ABOverlapDefinition}) with respect to the Gibbs measure:
\begin{eqnarray}
P_{AB}(q) = \<\delta\left( q-q_{AB}^{(1,2)}\right)\>\, .
\label{PAB_def}
\end{eqnarray}

In a liquid phase, $\<q^{(1,2)}_{AB}\>_{\rm state}$ vanishes and the  
function $P_{AB}(q)$ is a $\delta$-function.
In a glass phase $\<q^{(1,2)}_{AB}\>_{\rm state}> 0$ and  
the function $P_{AB}(q)$ 
becomes non-trivial,  with support in
the interval $[-\<q^{(1,2)}_{AB}\>_{\rm state},
\<q^{(1,2)}_{AB}\>_{\rm state}]$.  In the case of a continuous transition,
$\<F^{(1,2)}_d\>_{\rm state}$ and $\<q^{(1,2)}_{AB}\>_{\rm state}$ 
vanish at the transition point, while they exhibit a jump in the 
discontinuous case.

%
\section{Methods to study the glass phases in the cavity approach} 
\label{MethodsSection}

In this section we present the  methods that we  use to study the glass
transition on the Bethe lattice. They are applied to various types of polymers
in the next sections.
\subsection{Local instability towards a soft glass phase}
\label{LocalSection}

The simplest glass transition is the one associated with an instability of the 
liquid.
 The liquid solution is always embedded in the 1RSB
 formalism as the single pure state that exists at high 
temperature: it is described by  the field distribution 
$\rho({\bf p})=\delta({\bf p}-{\bf p}^*)$. This solution becomes locally unstable 
if fluctuations around ${\bf p}^*$ grow on average under the cavity recursion
(\ref{cavity0}-\ref{cavity2}).
This phenomenon occurs when
\be
	\label{instability}
	k\lambda_{\rm max}^2\ge 1
\ee
where $\lambda_{\rm max}$ is the largest eigenvalue of the transfer matrix for the propagation of deviations from the liquid under the recursion (\ref{cavity0})-(\ref{cavity2}), 
\be
	\mathcal{M}_{\alpha\alpha'}=\partial I_\alpha[{\bf p}^{(1)},\dots,{\bf p}^{(k)}]/\partial {p}^{(1)}_{\alpha'}|_{ {\bf p}^{(i)}={\bf p}^*}\,.
\label{Mdef}
\ee
(Notice that the stronger instability $k|\lambda_{\rm max}|=1$ \cite{MPSdeGennes} is irrelevant on a random
lattice, since it is associated to the establishment of a crystalline order
that is inherently frustrated because of the presence of
large loops.) Beyond the local instability, the distribution of local fields
$\rho({\bf p})$ becomes non-trivial, but it remains centered around the unstable
liquid fixed point. In physical terms this indicates that phase space begins
to divide up into a small number of states that comprise a large number of
microconfigurations. These states are characterized by weak local preferences
for certain polymer conformations that deviate only slightly from the
homogeneous liquid state.

The instability (\ref{instability})  generally develops below a temperature $T_i$.  Calling $T_{\rm cris}$ 
the temperature where the entropy vanishes, one can have two types of situations:
\begin{itemize}
\item When $T_i<T_{\rm cris}$, the local instability of the liquid  is clearly irrelevant, 
and a discontinuous
glass transition must take place   at some temperature  $ \ge T_{\rm cris}$. 
\item
 When $T_{\rm cris}<T_i$, either the instability drives a continuous glass transition
(as we will see in specific examples, this seems to be the generic case when the
instability occurs in a region where the liquid entropy is still large), or
there exists again a discontinuous glass transition taking place at 
temperatures $T>T_i$ and rendering the instability irrelevant. It is also possible,
that a first continuous glass transition towards a slightly frustrated phase
undergoes a successive discontinuous phase transition at lower temperatures
where a stronger degree of freezing takes place.
\end{itemize} 

Because of the relative simplicity of the liquid phase, it turns out that the
stability condition (\ref{instability}) can be studied explicitly for AB
copolymers with an interaction matrix which is symmetric under the 
$A\leftrightarrow B$ interchange. The detailed calculation is given in Appendix
\ref{Appendixinstab}. The dangerous eigenvalues $\lambda$ of the matrix $\cal M$ in
(\ref{Mdef}) are found to obey the equation 
\be 
\label{ABinstab}
\pm\frac{1}{k}\frac{w\sinh(\beta)}{1+w\cosh(\beta)}
=\frac{\lambda(1-(k\lambda)^{-L})}{(k-2)+k(k\lambda)^{-L}+2(k-1)
\sum_{i=1}^{L-1}q_i(k\lambda)^{-i}}
\ , 
\ee 
where the sign corresponds to ferromagnetic (+) and antiferromagnetic (-) 
interactions, respectively. The temperature dependent parameter 
$w=\sum_{a=1}^L p^*_{2a}/p^*_0$ characterizes the liquid solution 
and is independent of $L$, cf. App. \ref{LiquidApp} and Eqs. 
(\ref{liquidAB1}), (\ref{liquidAB2}). 
The sequence properties enter the above expression only through the 
autocorrelation function
$q_i=(1/L)\sum_{a=1}^L\sigma_a\sigma_{a+i}$.

The local instability
$\beta_i$ %
occurs at the smallest value of $\beta$ where the
characteristic polynomial (\ref{ABinstab}) has a root with $|\lambda|^2k=1$.
Usually, for attractive interactions between equal monomers, the relevant
eigenvalue is $\lambda=1/\sqrt{k}$ while the instability occurs in general 
with $\lambda=-1/\sqrt{k}$ in ampholytes.

The location of the instability for the various types of interactions and
sequences will be studied in the next sections. Let us just mention here that
the (periodic) Gaussian random bond heteropolymer generically undergoes a
discontinuous 1RSB glass transition, in agreement with previous studies \cite{ShakhnovichGutin89c}.

\subsection{Cavity recursion within the 1RSB approximation}
In order to study the glass phase itself, we need to compute
 the distribution of local fields of (\ref{rho(p)}) for the Bethe lattice.
We shall do it here within the 1RSB cavity formalism of (\cite{cavity0,cavityT}).
We shall not rederive the full formalism but give the main ingredients
needed for our study.
The statistical average of the simple cavity recursion 
(\ref{cavity0}-\ref{cavity2}), which holds within a given pure state,  
leads to a recursion 
relation for this distribution:
\begin{eqnarray}
\label{onestepcavity}
\rho({\bf p})=\frac{1}{\mathcal{Z}}\int \prod_{i=1}^k d\rho({\bf p}^{(i)}) \;
\delta({\bf p}-I[{\bf p}^{(1)},\dots,{\bf p}^{(k)}])\; e^{-m\beta\Delta f[{\bf p}^{(1)},
\dots,{\bf p}^{(k)}]}
\end{eqnarray}
where $I[{\bf p}^{(1)},\dots,{\bf p}^{(k)}]$ is given by (\ref{cavity0})-(\ref{cavity2}), 
and $\mathcal{Z}$ is a normalization. The non trivial reweighting,
which depends on the parameter $m$ defined in Section~\ref{sect:mdef},
involves  the free energy change induced by the recursion,
which is given by
$\Delta f[\{{\bf p}^{(i)}\}]\equiv -\beta^{-1}\log( C[\{ {\bf p}^{(i)} \}])$ ,
where $C[\{ {\bf p}^{(i)} \}]$ is the normalization term appearing in  (\ref{cavity0})-(\ref{cavity2}). This reweighting
accounts for
the fact
that the number of pure states increases exponentially with their free 
energy.

The  free energy is obtained by properly weighting the
contributions of different pure states:
\begin{eqnarray}
\label{onestepphi}
\beta m\phi_{1}(m)&=&-\log\left[\int\prod_{i=1}^{k+1}\!
d\rho({\bf p}^{(i)})\, w^m_s(\{{\bf p}^{(i)}\})\right]
+\frac{k+1}{2}\log\left[\int\prod_{i=1}^{2}d\rho({\bf p}^{(i)})\, 
w^m_l({\bf p}^{(1)},{\bf p}^{(2)})\right]\, ,
\end{eqnarray}
where $w_s$ and $w_l$ are the site and link partition functions
defined in Eqs.~(\ref{SitePartitionFunction}) and 
(\ref{LinkPartitionFunction}). The complexity $\Sigma(\omega)$
is obtained from $\phi_1(m)$ through a Legendre transform: 
$m\beta\phi_1(m)=m\beta \omega-\Sigma(\omega)$.
Note that the recursion relation (\ref{onestepcavity}) is the 
saddle point equation for the functional $\phi_{1}(m)$ with 
respect to $\rho({\bf p})$.

Close to a continuous glass transition, $\rho$ is strongly peaked around the liquid fixed point ${\bf p}^*$, and we can expand the free energy as a function of the moments of the fluctuations ${\bf p}-{\bf p}^*$ over the pure states, as outlined in Appendix \ref{Appendixmoments}. To leading order the corrections to the liquid free energy arise from fluctuations in the ``replicon'' mode, the unstable direction of the transfer matrix (\ref{Mdef}), whose magnitude grows as $(T_i-T)^{1/2}$. The continuous glass transition is found to be of third order,

\be 
	\label{1orderphi}
	\phi_1-\phi_{\rm liq}= c \frac{1-m}{(2-m)^2} (T_i-T)^3+ O\left((T_i-T)^4\right)
\ee
where $c$ is a positive constant. This is in contrast to discontinuous 
glass transitions which are (generally) of second order in the free energy. 
%

\section{Two exemplary cases: The alternating ampholyte and HP model}
\label{alternatingsequences}

In this section we apply the cavity 1RSB formalism to two specific
sequences: the regularly alternating copolymer chains $ABABAB\dots$ for
ampholytic and symmetrized-HP interactions. These turn out to be rather extreme
representatives in the ensemble of possible neutral copolymers, but they
are the simplest ones, and they
exhibit the characteristics of the continuous (ampholyte) and discontinuous
(HP) transition in a very clear manner.

The folding of an alternating copolymer on a regular Bethe lattice is a
frustrated problem, while, clearly, on a regular cubic lattice, it would just
behave as a homopolymer with homogeneous interactions $E_{AB}\equiv e$.  
However, we expect that as soon as a certain number of defects are
introduced in such sequences, their folding on the cubic lattice will be
similarly frustrated. In terms of Markovian sequences, we consider here the case of
$\pi\ll 1$. 

While these sequences are expected 
to behave differently from the alternating  one $\pi=0$ on the cubic lattice,
 it is reasonable 
to assume that the  $\pi \to 0$ limit is smooth on the Bethe lattice.
Then the Bethe approximation of $\pi\ll 1$ sequences can be studied
using the perfectly alternating sequence, as we do here here. 
Alternating chains are more easily studied with the cavity
method, since the number of local fields may be
reduced to 5 (with 4 independent degrees of freedom): due to the inversion
symmetry the local conformations reduce to $\alpha\in\{0,1A,1B,2A,2B\}$, where
$1A$ ($1B$) comprises the two conformations $\ua A$ and $\da A$
($\ua B$ and $\da B$). 
The cavity recursion relation (\ref{onestepcavity}) can thus be handled
relatively easily, using a population dynamics algorithm described in App.
\ref{app:popdyn}

\subsection{Ordered structures, correlations, frustration and the order of 
the glass transition} 
Before embarking on the details of the cavity computation for the alternating
chains, we present here some simple arguments
explaining the very different physical nature of the glass phase
in  the alternating ampholyte, which has a continuous
transition, and in the symmetrized HP model, which has a discontinuous transition.

Instead of a Bethe lattice, let us consider a regular tree and ask for a maximally dense polymer configuration such that all interactions are satisfied ($AB$ interactions in ampholytes and $AA$ or $BB$ interactions in the symmetrized HP model). In Fig.~\ref{order} we show  typical configurations for each case. 
\begin{figure}
	\resizebox{9 cm}{!}{
  \includegraphics{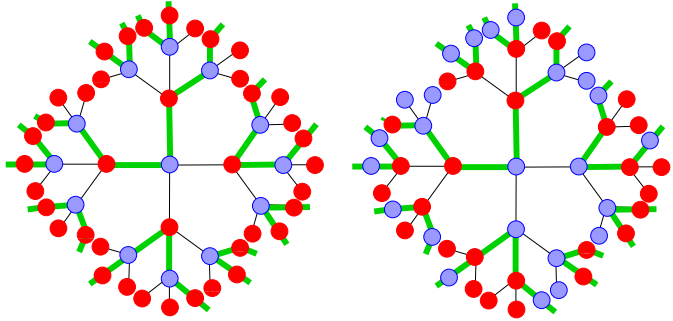}}
\caption{Unfrustrated, maximally dense structures on a tree ($k=3$). 
The ampholyte (left) has an evident stratified order and long range correlations of site occupancies. The location of $A$'s and $B$'s in the HP-polymer is correlated with the backbone configuration (thick edges) which makes the distribution of monomers look random.}
\label{order}
\end{figure}
While there is a stratified order in 
ampholytic configurations
that manifests itself in strong long range correlations, 
the symmetrized HP model has an ``ordered''
structure that is highly correlated with the backbone configuration. No long
range correlations may persist, and this dense ground state
is difficult to distinguish from a dense liquid configuration.

If we turn back to a Bethe lattice, frustration is induced by the presence of
large loops. Odd loops are inherently frustrated in the ampholyte since they
necessarily have to break up the long range correlations of the layered
structure. This is not the case in the HP-like model where most constraints from
loops can be satisfied when the backbone is arranged in the right way. In
other words, the information about local conformations and the associated
constraints cannot be propagated far away in the case of the HP-like 
chain, since the
correlations of ordered structures die out quickly with distance. As long as
the density is not too large and there are sufficient voids in the globule, a
global frustration will not be able to establish. For the ampholyte, however,
it will be favorable, even at lower density, to develop local (site)
preferences for a certain monomer type and thus increase the probability of
satisfied interactions. This mechanism is at the basis of the instability of
local fields in the liquid. Note that in the first place this instability is
related to the type of monomer accommodated on a given site rather than the
backbone structure. The latter will only come into play at larger
densities/lower temperatures.

This qualitative discussion applies equally to correlated sequences which
are not perfectly alternating but have a strong tendency to alternate (small $\pi$).
At the other extreme, if one considers the case
of $\pi$ close to one, where consecutive monomers tend to be alike,
one can apply the same type of considerations, 
but with the roles of ampholyte and HP-like chain
reversed. We can thus conclude  that the local instability
of a HP-like chain with long blocks of like monomers is associated
to the appearance of pure states characterized by the \textit{same} monomer
preferences for small regions on the lattice. This is reminiscent
of the
microphase separation (MPS) \cite{Microphaseseparation} which has 
been much discussed in this context and becomes relevant for sequences with a 
distinct block structure
\cite{SfatosGutin93,GutinSfatos94,Dobrynin95,Heteropolymerreview}. 
However one should remember that the present formulation of the cavity method, 
which neglects small loops in the lattice, does not allow any quantitative
study of this phenomenon (this could be addressed using more refined cluster 
variational methods).

Repeating the above arguments for more general cases of short range 
correlated sequences, one sees that in general a local instability is 
favored by sequences whose monomer distribution tends to be annealed 
(e.g., ampholytes with a tendency towards charge alternation along the 
sequence). It is interesting to note that such `annealed sequences'
 naturally result from common protein design 
schemes~\cite{Sequencedesign1,Sequencedesign2, Copolymerdesign, LevyFlight}. 
 
\subsection{The continuous transition in the AB ampholyte}

We start our quantitative study with the alternating ampholyte 
on a lattice with connectivity $k+1=6$. 
For this polymer, the local instability of the liquid found from 
(\ref{ABinstab})
develops at the inverse temperature $\beta_i\approx 0.7947$, 
much smaller than in most other neutral
sequences.
The Parisi parameter $m$ remains small throughout this phase.

A closer analysis of the instability shows that the most
unstable eigenvector is
antisymmetric with respect to the exchange of $A$ and $B$. This indicates that
the pure states are essentially characterized by the preference of the sites
to accommodate one of the two monomer species, in agreement
with our qualitative discussion. 

On lowering the temperature, the preference of sites for 
certain conformations (and not only for the respective monomers), increases.
This could be interpreted  as a growing degree of freezing that affects 
larger and larger length scales. 

Figure \ref{Fig_AF} shows the basic thermodynamic observables $\rho,s,u$
in the glass phase, computed in the cavity method 
 and compares them to the values
found in the unstable liquid solution. The data have been computed on the
coexistence line, i.e., by fixing $\mu_c(\beta)$ such that the glass static
free energy vanishes, $\phi_1(m_s;\mu_c)=0$ (as explained in App. \ref{ap:coex}).
The strong frustration of the polymer can clearly be seen from the suppression
of the density in the glass phase that saturates around $\rho=0.71$, while in
a liquid phase it would tend to $\rho=1$. The entropy crisis of the liquid is
prevented, the internal entropy of the pure states remaining rather large even
at low temperature. There is no sign of a  strong (discontinuous) 
freezing transition.
\begin{figure}
	\resizebox{9 cm}{!}{
  \includegraphics{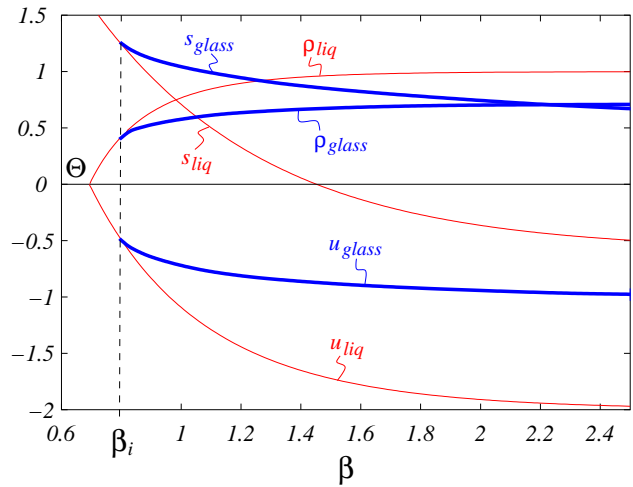}}
\caption{Alternating ampholyte on a lattice with 6 neighbors per site. 
The thick lines show the density $\rho$, entropy $s$ and internal energy $u$
computed in the glass phase  using the 1RSB approximation. The thin lines
give the corresponding values in the liquid solution, which is unstable
beyond $\beta_i\approx 0.7947$. The glass transition is continuous.}
\label{Fig_AF}
\end{figure}

In App. \ref{app:op} we explain how to compute the order parameter (\ref{orpar_space}) within the cavity approximation. The result for the alternating ampholyte is shown in Fig.~\ref{conffreezeA}, which again shows a
continuous transition.
\begin{figure}
	\resizebox{9 cm}{!}{
  \includegraphics{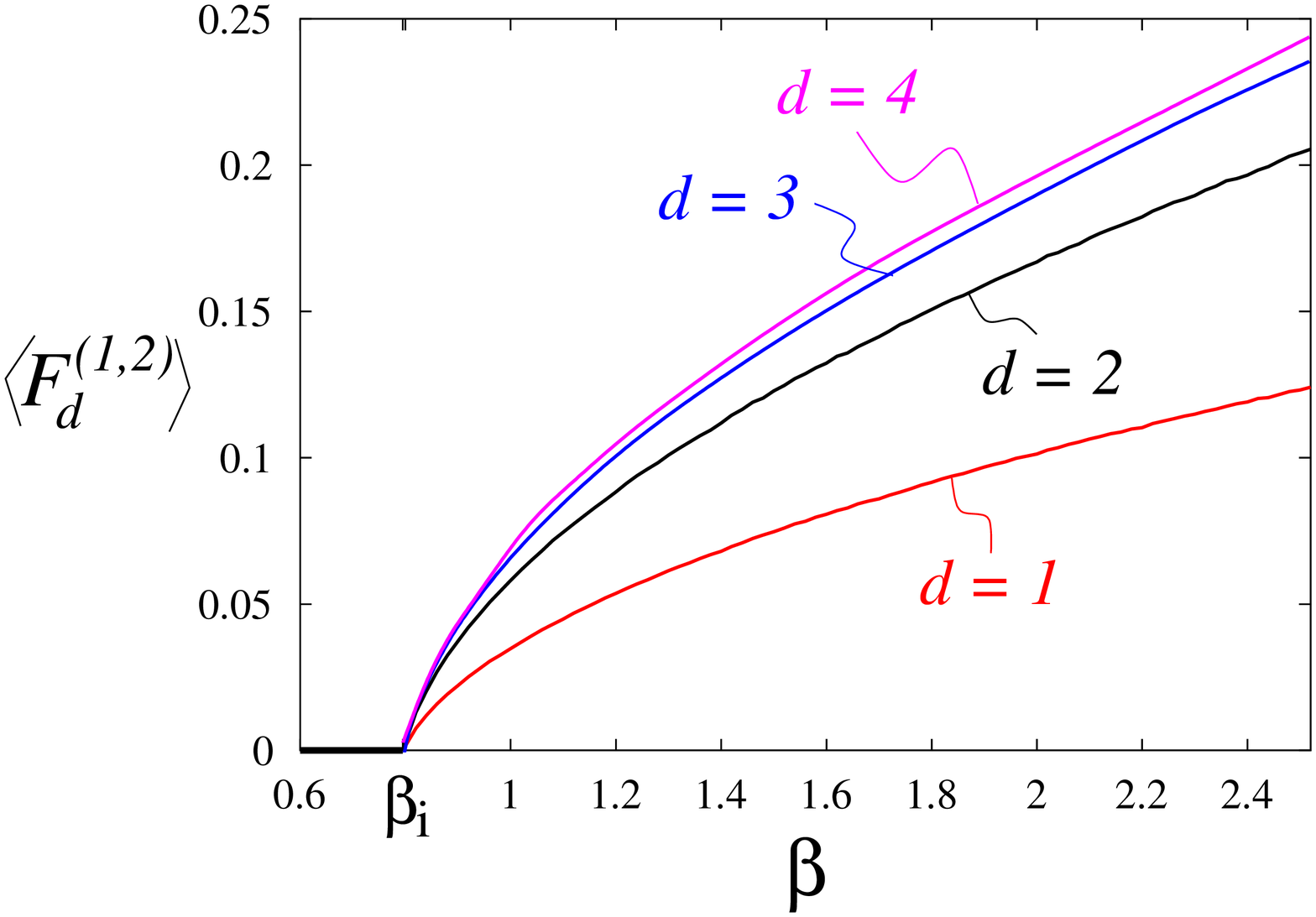}}
	\caption{Alternating ampholyte: The order parameters 
$\< F_d^{(1,2)}\>_{\rm state}$ for the glass phase,
defined as the time persistent part of the distance vector between monomers at a distance $d=1,2,3,4$ in the backbone (see (\ref{orpar_space})), 
plotted versus the inverse temperature $\beta$.}
\label{conffreezeA}	
\end{figure}

\subsection{The discontinuous transition in the alternating HP model}
\label{AlternatingHP}

The case of the symmetrized-HP alternating sequence, 
always on a lattice with connectivity $k+1=6$,
 is extreme in the opposite sense.
The liquid solution is always locally stable,  even in the region of negative
entropy. However, running the population dynamics algorithm
for the 1RSB cavity method, one finds a discontinuous glass transition.
The dynamic transition takes place at $\beta_d\approx
1.387$,
just before the entropy crisis of the liquid
($\beta_{\rm cris}=1.4525$).  The static phase transition follows at
$\beta_s\approx 1.442$, in a region of very high density, $\rho\approx 0.95$,
and almost vanishing entropy. In Fig.~\ref{Fig_F}, we plot the density, entropy
and internal energy for the  alternating HP-polymer along the coexistence curve. 
The internal entropy
of the statically dominating pure states is seen to nearly vanish in the
frozen phase, and the system barely evolves upon lowering the temperature.
This scenario is very similar to the abrupt freezing encountered in the random
energy model (REM).

\begin{figure}
	\resizebox{9 cm}{!}{
  \includegraphics{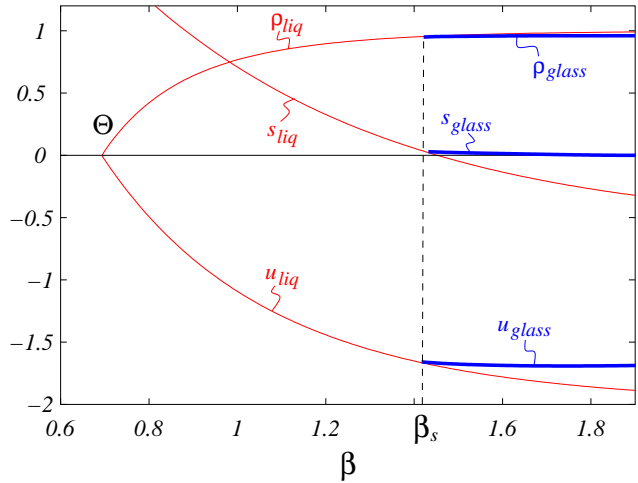}}
\caption{
Alternating HP-like polymer on a lattice with 6 neighbors per site: 
The thick lines show the density $\rho$, entropy $s$ and internal energy $u$
computed in the glass phase using the 1RSB approximation. The thin lines
give the corresponding values in the liquid solution, which is always
locally stable. The glass transition is a discontinuous one;
it is an almost perfect freezing transition as in the REM.}
\label{Fig_F}
\end{figure}

The  computation of  the  order parameter  (\ref{orpar_space}) proceeds
as in the case of the ampholyte. 
The result is shown in Fig.~\ref{conffreezeF} and shows clearly the
discontinuous transition.
\begin{figure}
	\resizebox{9 cm}{!}{
  \includegraphics{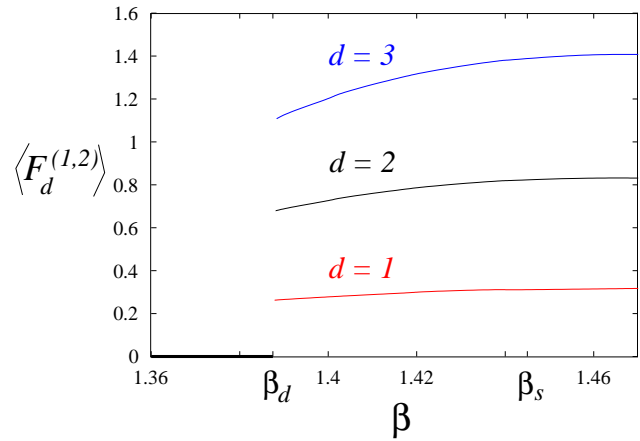}}
	\caption{Alternating HP-like polymer: The structural order parameters $\< F_d^{(1,2)} \>_{\rm state}$ for the glass phase, plotted versus the inverse temperature $\beta$. Note the considerably higher values than in Fig.~\ref{conffreezeA}, indicating a much stronger freezing of local conformational degrees of freedom.}\label{conffreezeF}	
\end{figure}

%
%
\subsection{Numerical simulations}
\label{NumericalBethe}

As we already stressed, one advantage of our approach consists 
in the possibility of checking mean field computations 
using numerical simulations of
well defined polymer models on a Bethe lattice. 
Here we want to demonstrate this feature by 
considering the alternating AB ampholyte.
\begin{figure} 
\centerline{
\epsfig{figure=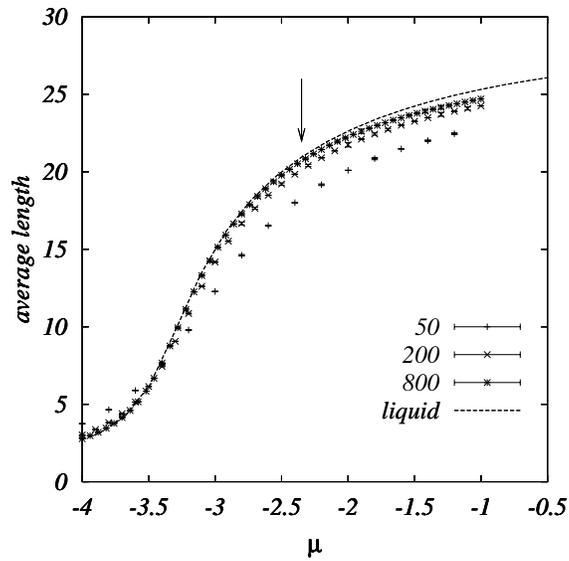,width=0.5\linewidth}}
\caption{Average length of the polymers simulated on the Bethe lattice. 
Sizes of 
the lattice are indicated in the legend. The arrow signal the
liquid-soft glass phase transition.}
\label{LengthNumerical}
\end{figure}

We made extensive simulations on Bethe lattices with connectivity 
$(k+1)=6$ and volumes $V$ ranging from $100$ to $800$. 
For all of the data presented in this Section, we fixed 
$\beta=2.0$ above
the $\Theta$-point inverse temperature $\beta_{\Theta} \approx 0.693$
and varied the chemical potential $\mu$. As $\mu$ is increased, the 
system undergoes at first a second order collapse transition 
(at $\mu\approx -3.21887$) and then a continuous glass transition
to the soft-glass phase ($\mu_i(\beta)\approx -2.38431$).

Notice that  most of the algorithms for simulating polymers on 
finite-dimensional graphs cannot be applied to the Bethe lattice.
In fact local moves are impossible because of the absence of short
loops. On the other hand, global moves would require a detailed
knowledge of the loop structure for any graph realization. 

This problem can be overcome by simulating a melt of variable-length 
polymers, the length being finite in the thermodynamic limit. 
The single-polymer physics is recovered when the average 
length diverges.
We refer to App. \ref{app:mcB} for  a detailed description of our algorithm.
In Fig. \ref{LengthNumerical} we show our numerical data for the average 
polymer length $\<l\>$. Notice that $\<l\> \approx 10\div 25$ within the 
dense phase. As will be clear from the other numerical results, this is enough
for assuring small deviations from the infinite-length limit.  
The main effects are: a rounding of the collapse transition
and a small shift of the soft glass transition (which occurs at 
$\mu_i(\beta, \mbox{finite } l)\approx -2.40923$).

In order to achieve equilibration within the soft glass phase 
we adopted the parallel tempering technique 
\cite{Nemoto96,Marinari98}. We
tested equilibration using the method of Ref. \cite{BhattYoung88}, 
and always checked 
the acceptance rate for temperature-exchange moves to be larger than $50\%$.

\begin{figure}
\begin{tabular}{cc}
\epsfig{figure=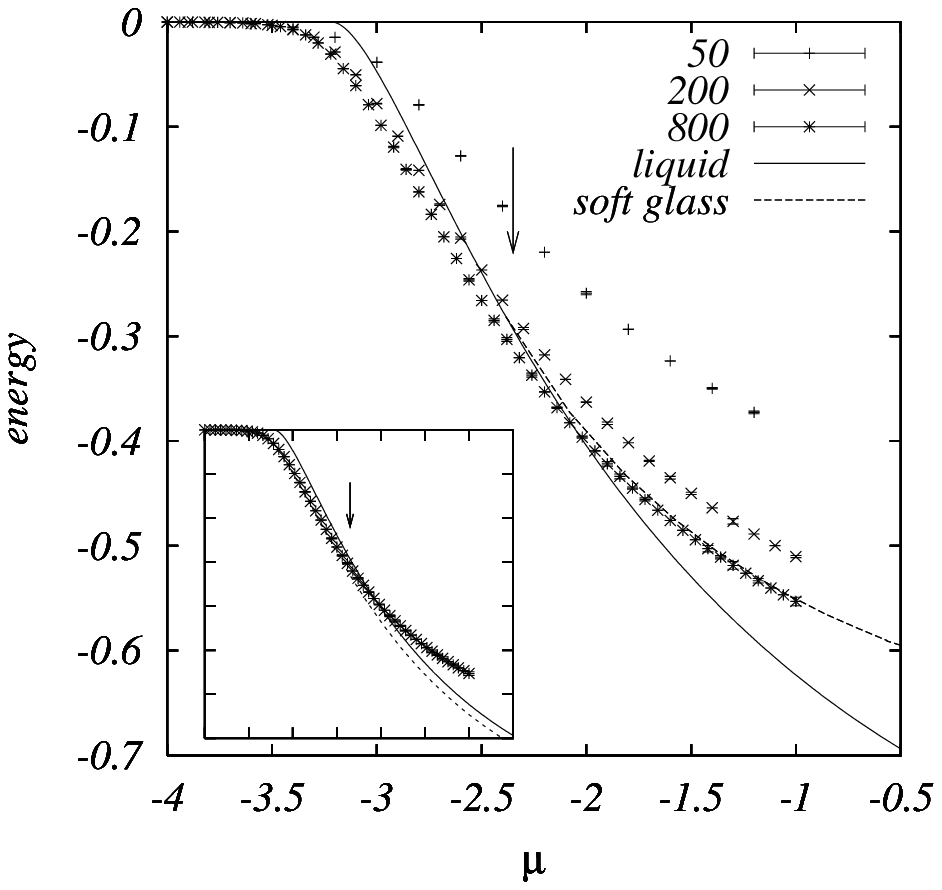,width=0.5\linewidth} &
\epsfig{figure=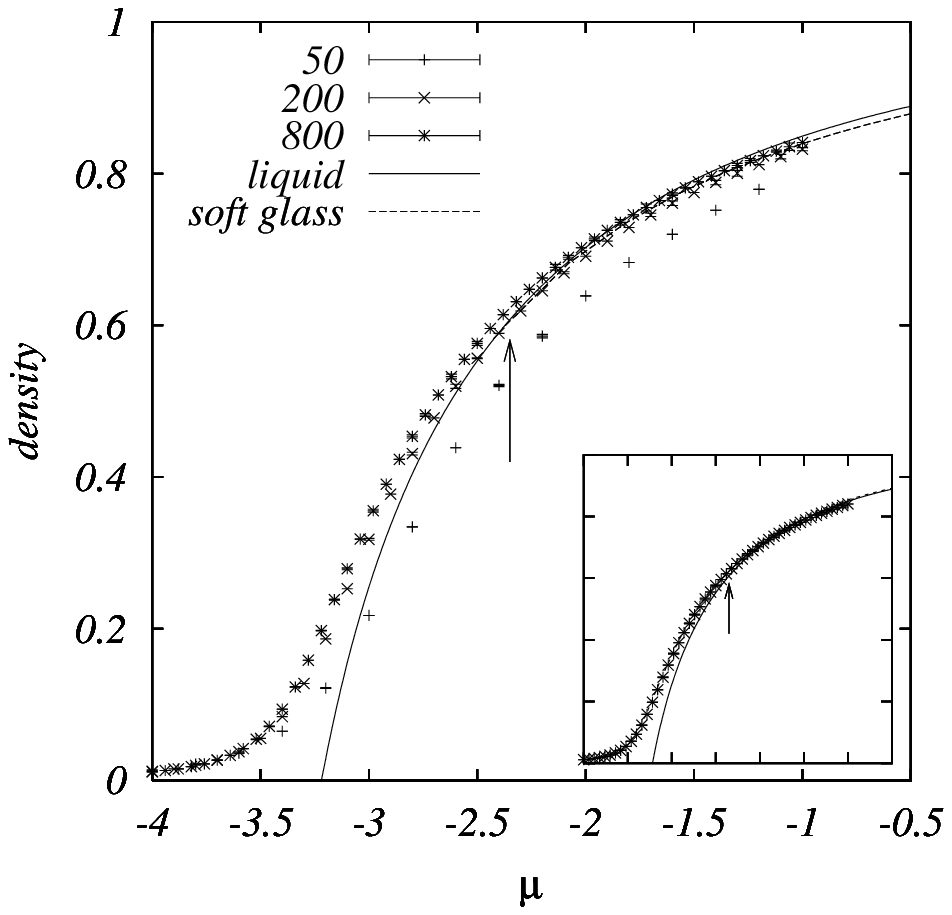,width=0.5\linewidth}
\end{tabular}
\caption{Simulations of the alternating ampholyte on the Bethe lattice
with connectivity $k+1=6$. 
The energy per site of the polymer (left) and its density (right)
are plotted versus the chemical potential.
In the main frames numerical data (symbols) for various lattice sizes
 are compared with the cavity results (dashed line)
for average polymer length $\<l\> \to \infty$. 
The agreement is very good. In the insets we plot
the liquid prediction for infinite (continuous line) and finite 
(dashed line) average polymer length, which shows that the 
finite length corrections are already small. Notice that 
in the density inset the finite-length theoretical curve
is barely visible because it is superimposed on the data.
The arrows indicate the analytic
result for the glass transition point $\mu_i$. }
\label{EnergyDensityNumerical}
\end{figure}
In Fig.~\ref{EnergyDensityNumerical} we plot the energy per lattice site and
the monomer density, as functions of the chemical potential $\mu$. Notice 
that the liquid - soft
glass phase transition is barely discernible from the monomer density, and the
energy curve is also quite smooth. The 1RSB cavity result gives a very good
quantitative description of the transition.

\begin{figure}
\begin{tabular}{cc}
\epsfig{figure=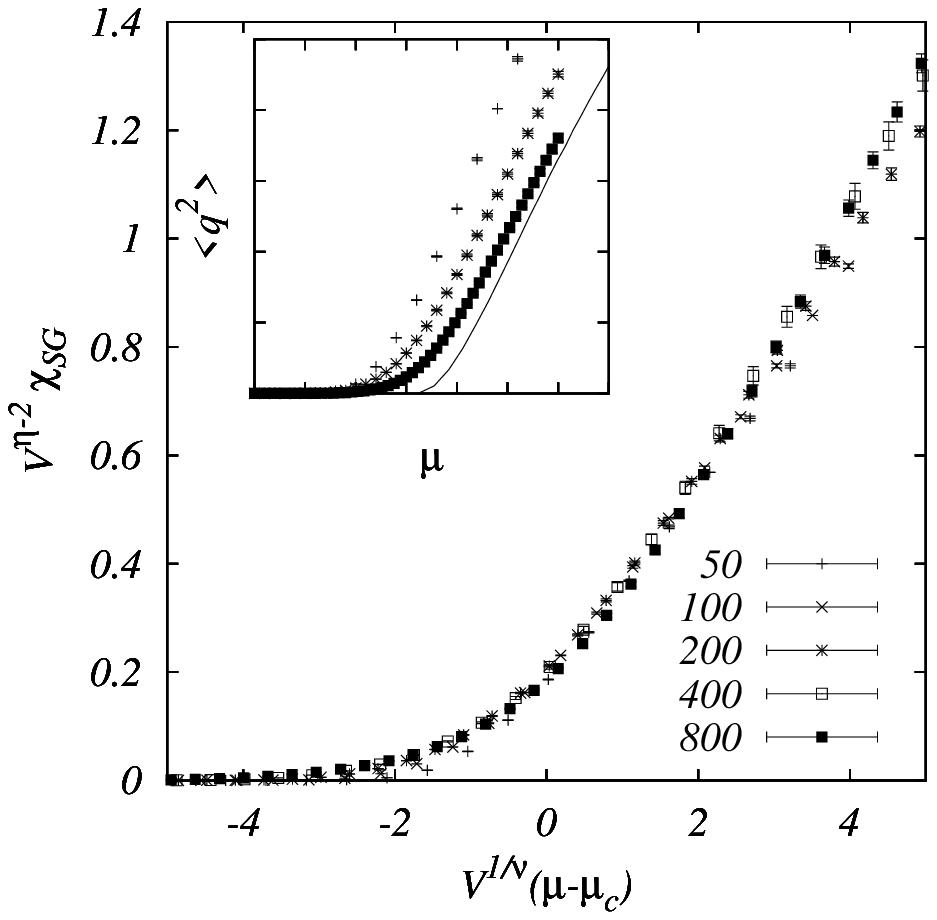,width=0.5\linewidth}&
\epsfig{figure=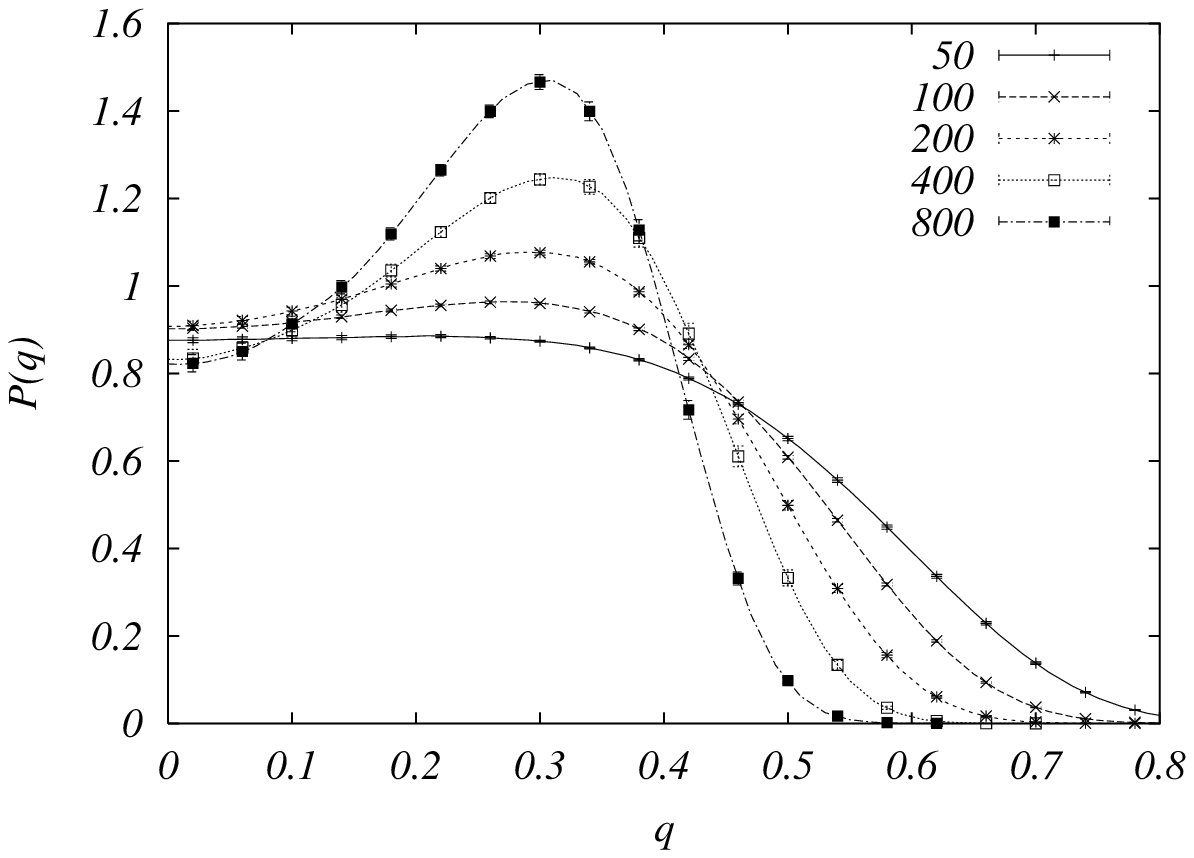,width=0.5\linewidth}
\end{tabular}
\caption{Left: Finite-size scaling of the spin glass susceptibility,
that exhibits a clear divergence as a function of system size at the 
expected value $\mu_c$. Right: probability distribution of the AB overlap, 
evaluated in the soft glass phase at $\mu = -1$.
In the inset of the left frame we compare the second moment of the AB
overlap $\overline{\<q_{AB}^2\>}$ with the analytical prediction.}
\label{OverlapNumerical}
\end{figure}
In order to get a finer description of the glass phase, we
 have measured the order parameter function 
$P_{AB}(q)$ defined in (\ref{PAB_def}). In 
Fig. \ref{OverlapNumerical} we report our numerical data
for this quantity at the highest chemical potential considered ($\mu=-1$).
Because of the large finite-$V$ effects, it would be difficult
to conclude from the numerics alone that 
the infinite-$V$ function is non-trivial. However, the
data agree with the 1RSB predictions for  
the Edwards-Anderson parameter, $q_{\rm EA}\approx 0.259$.

In the same figure  (left frame) we consider the spin-glass susceptibility:
\begin{eqnarray}
\chi_{\rm SG} =
\frac{1}{V} \sum_{i,j}\overline{\<s_i s_j\>^2}
= V\overline{\<\left[q_{AB}(s^{(1)},s^{(2)})\right]^2\>}  \, ,
\end{eqnarray}
This quantity diverges as $\mu\to \mu_c^-$ in the thermodynamic limit. 
In a finite size sample, its behavior is ruled by the usual finite-size 
scaling form
\begin{eqnarray}
\chi_{\rm SG}(V,\mu) = V^{2-\eta}\, \overline{\chi}[V^{1/\nu}(\mu-\mu_c)]\, .
\label{FSSForm}
\end{eqnarray}
From the cavity solution of the model, one finds that
$\overline{\<q_{AB}^2\>} \approx A(\beta) [\mu - \mu_c(\beta)]^2$
for $\mu\gtrsim\mu_c(\beta)$.
This result implies the following relation between the critical exponents
defined in Eq. (\ref{FSSForm}): 
\begin{eqnarray}
2-\eta+2/\nu = 1\, .
\end{eqnarray}
In fact we find a nice collapse of data corresponding to different sizes
using $\nu = 4$ and $\eta = 3/2$. The comparison of $\<q_{AB}^2\>$
with the 1RSB cavity prediction is quite good.

\begin{figure}
\centerline{
\epsfig{figure=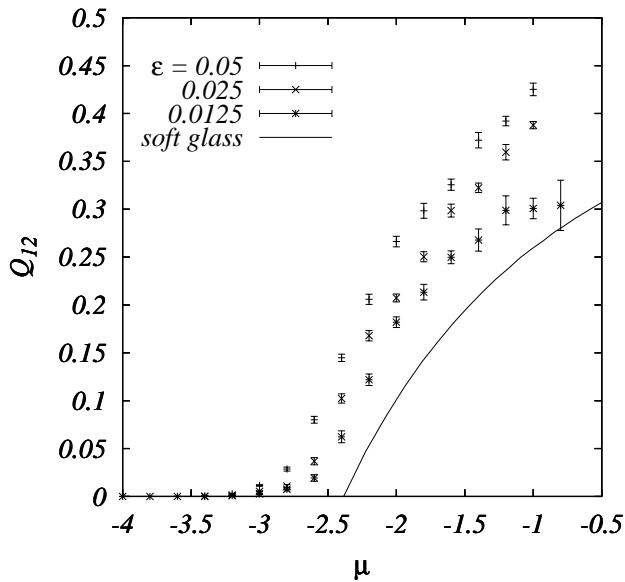,width=0.55\linewidth}}
\caption{Average AB overlap $\overline{\< q_{AB}\>}$ among two replicas
coupled through a term of the type $-N\epsilon q_{\rm AB}$. The full line
is the prediction from the cavity method for the zero $\epsilon$ limit.
Here $V=10^4$.}
\label{EpsilonCouplingNumerical}
\end{figure}
An alternative approach for exploring the low energy structure of
the system consists
 in coupling two replicas through their overlap,
cf. Eq. (\ref{ABOverlapDefinition}). In practice, one adds a term of
the form $-N\beta\epsilon q_{AB}(s^{(1)},s^{(2)})$ to the two-replica 
Hamiltonian and tries to estimate $q_{\rm EA}$ as follows
\begin{eqnarray}
q_{\rm EA} = \lim_{\epsilon\to 0}\lim_{N\to\infty} 
\overline{\<q_{\rm AB}(s^{(1)},s^{(2)})\>_{N,\epsilon}}\, .
\end{eqnarray}
In Fig. \ref{EpsilonCouplingNumerical} we show the numerical results for
$\overline{\<q_{\rm AB}\>_{N,\epsilon}}$ on a large size lattice
($V=10^4$) and several values of $\epsilon$. In order to 
simulate large lattices, we did not use parallel tempering here. Furthermore, we adopted a weaker equilibration criterium, requiring 
$\overline{\<q_{\rm AB}\>_{N,\epsilon}}$ to be roughly time-independent on a 
logarithmic scale.
Once again, the numerical results compare favorably with the outcome
of the cavity calculation.
%
%
\section{Random Markovian copolymers}
\label{MarkovianSection}

One can show using the formula (\ref{ABinstab}) that the local 
instability appears the earlier, the stronger the tendency of monomers 
to be annealed along the sequence, that is, the more $A$'s and $B$'s 
tend to alternate in an ampholyte, or to form blocks in an HP model. 
In both cases the autocorrelation function $q_i$ is large and its sign 
oscillates  (alternating sequence) or remain positive (`blocky' sequence).

To be more quantitative, let us consider a random copolymer 
chain in the limit $L\ra \infty$
characterized by the probability $\pi \in [0,1]$ of two neighboring 
monomers to be of the same type. The autocorrelation function of such 
a chain is (in the $L\to\infty$ limit) $q_i=(2\pi-1)^i$.    

In Figs.~\ref{MarkovchainAF} and \ref{MarkovchainF} we plot the 
inverse temperature $\beta_i$ at the local instability as a function of the 
parameter $\pi$ for the ampholyte and symmetrized-HP models. 

This instability is certainly 
irrelevant when $\beta_i$ is larger than the inverse temperature
of the entropy crisis of the liquid, 
$\beta_{\rm cris} = 1.4525$. This situation occurs for $ \pi>0.4480$ in ampholytes,
and proves the existence of a discontinuous transition.
But already when $\beta_i$ is smaller than, but close to, $\beta_{\rm cris}$,
one should expect a discontinuous 1RSB  transition to take place
at a $\beta<\beta_i$.

In order to complete the diagram, we have numerically solved the 
cavity recursion by population dynamics for neutral sequences of 
period $L=20$, but otherwise random composition. From the experience 
gained for the extreme case of the alternating HP-model (see below), 
we expected a kind of frozen solution with rather strong local conformational 
preferences to dominate the low temperature phase. Such a solution is 
rather non-trivial to find in a huge functional space, in particular 
since it has to be expected that it occurs in a discontinuous manner 
and cannot in general be found by randomly perturbing the liquid solution.

We therefore proceeded by initializing the population in a highly 
polarized state that we will discuss in more detail in the next Section. 
This state actually corresponds to an unstable fixed point, but it turns 
out that at low temperatures, it is usually quite close to a stable 
non-trivial solution of the 1RSB cavity equations. At each temperature, 
we iterated the cavity recursion for about 100 sweeps of the population 
dynamics, cf. App. \ref{app:popdyn},
fixing the chemical potential to its liquid critical value, since  this 
value describes correctly the thermodynamic equilibrium up to the static 
phase transition. The Parisi 
reweighting parameter was set to $m=1$ in order to detect the dynamic 
transition, i.e., the local instability of the frozen solution. For reasons 
of numerical stability, we restricted ourselves to sequences with an 
anti-palindromic structure, i.e., sequences invariant under inversion 
and subsequent exchange of $A$'s and $B$'s. The field distributions 
$\rho({\bf p})$ inherit this invariance, and thus in each update of a new 
cavity field we can decide at random to apply a symmetry operation 
to the new fields first. This stabilizes the iteration since 
it counteracts the numerical tendency to spontaneously break the balance 
between $\ua$- and $\da$-states. Indeed, there is a gauge degree of freedom 
associated to the relative weight of the two orientations of the chain, 
and in general it is difficult to maintain them balanced, while it can be 
enforced in sequences with a palindromic symmetry. The reason to choose 
\textit{ anti}palindromic rather than palindromic ones is to avoid at the 
same time an asymmetry between $A$- and $B$-states which likely occurs in 
small populations, in particular in the case of attractive 
interactions among equal monomers.

Our findings for the sequences of period $L=20$ are summarized in the 
plots~\ref{MarkovchainAF}, \ref{MarkovchainF} and \ref{variances}. 
Figure \ref{variances} shows the variance (square of the standard deviation) 
of the local field for
$\ua (a=1)$ over the distribution $\rho({\bf p})$ for several sequences as a 
function of inverse temperature. This is a measure for the degree of the 
local bias away from the liquid. Almost independently of the particular 
sequence statistics we find that for $\beta>\beta_d \approx 1.23$ a 
strongly frozen phase (with very low internal entropy) exists with an 
associated dynamic transition at $\beta_d$. Depending on the sequence 
statistics, the regime of higher temperatures is either entirely liquid 
(e.g., for $\pi \le 0.50$ in the ampholytes), or exhibits a weaker form 
of frustration in a phase of presumably fully broken replica symmetry. 
The latter continuously joins the liquid solution at the local instability 
predicted by (\ref{ABinstab}). For the phase diagram in the $\beta-\mu$-plane for either of the two scenarios we refer to Figs.~\ref{phasedia}.

\begin{figure}
	\resizebox{9 cm}{!}{
  \includegraphics{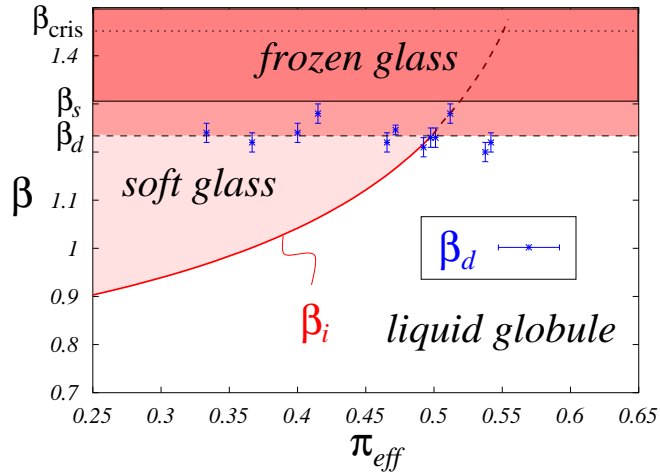}}
	\caption{Phase diagram for ampholytes as a function of sequence correlations and inverse temperature. The continuous line indicates the local instability $\beta_i$ of the liquid as a function of the Markov chain parameter $\pi$. The points with error bars indicate the dynamic phase transitions $\beta_d$ found numerically for several sequences of period $L=20$ whereby we associated an effective parameter $\pi_{\rm eff}$ to each chain such that the local instability predicted from $\pi_{\rm eff}$ coincides with the actual one. Almost independently of the chain composition we find a highly frozen phase beyond $\beta_d\approx 1.23$ that is reached via a discontinuous glass transition well before the liquid would undergo an entropy crisis at $\beta_{\rm cris}$. For $\pi\leq 0.50$, this freezing is preceded by a continuous glass transition, as predicted from the local stability analysis of the liquid. The actual thermodynamic freezing transition occurs at a lower temperature $\beta_s>\beta_d$. The horizontal lines for the static and dynamic transitions are an educated guess for the location of these transitions in the limit $L\ra \infty$.}
\label{MarkovchainAF}
\end{figure}

\begin{figure}
	\resizebox{9 cm}{!}{
  \includegraphics{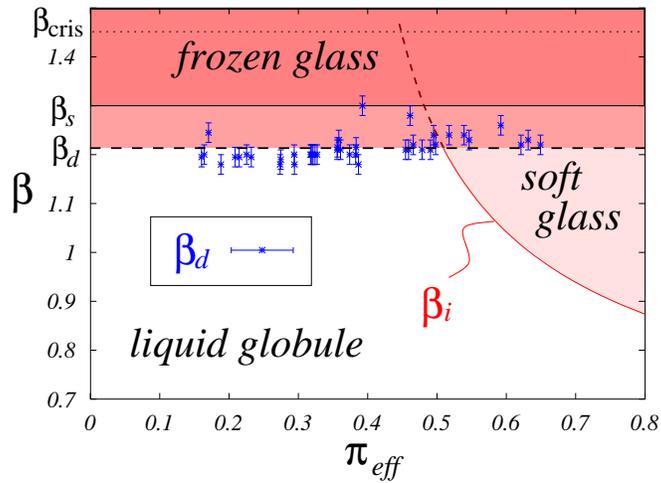}}
	\caption{
The same general phenomenology as in the ampholyte, cf. 
Fig. \ref{MarkovchainAF}, holds for the HP-type models, but here, 
the continuous transition takes place at $\pi\geq 0.5$. 
Notice that the $\pi_{\rm eff}$ window displayed here is larger than in Fig. \ref{MarkovchainAF}.} 
\label{MarkovchainF}
\end{figure}

\begin{figure}
	\resizebox{9 cm}{!}{
  \includegraphics{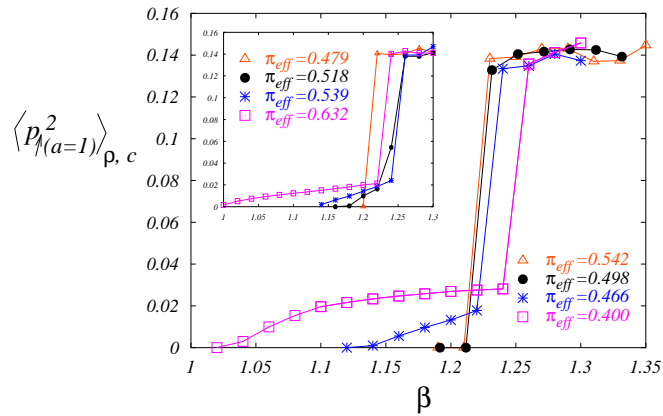}}
	\caption{The variance $\<p_{\ua (a=1)}^2\>_\rho-\<p_{\ua (a=1)}\>_\rho^2$ of a selected local field as a function of inverse temperature for a variety of anti-palindromic sequences of period $L=20$. In general, there is a very distinct discontinuous transition around $\beta_d\approx 1.23$, that is preceded by a glassy regime with smaller fluctuations in the local fields if the sequence has a tendency for anticorrelation in ampholytes (main frame) or correlation in symmetrized HP-like chains (inset). The sequences are characterized by their effective Markov chain parameter 
$\pi_{\rm eff}$
as in Figs.~\ref{MarkovchainAF} and \ref{MarkovchainF}.} 
\label{variances}
\end{figure}

The generic picture of a quench in temperature is thus the following: For ampholyte sequences with some tendency to alternation or HP-like-sequences with a preference for block formation, there is a continuous glass transition whose location depends strongly on the composition of the sequence. The corresponding glass phase is characterized by a relatively weak frustration and a rather small number of states that comprise many microconfigurations with some weak local preferences for certain conformations.
This preliminary glass phase undergoes a further discontinuous phase transition at a lower temperature $\beta_d\approx 1.23$ that is almost independent of the sequence structure and might be called the effective freezing transition. For sequences with correlations of the opposite kind, the freezing transition is the only phase transition and occurs directly from the liquid.
It is interesting to note that in numerical simulations of the folding dynamics of neutral HP-type copolymers, the dynamical glass transition was also found to be essentially independent of the sequence \cite{BryngelsonReview95}. 

It is intriguing that the critical parameter of $\pi$ separating the FRSB from the 1RSB freezing scenario is very close to $\pi=1/2$ which corresponds to sequences without correlations. This is particularly interesting from the point of view of protein folding. The nature of correlations present in the amino acid sequences of natural proteins is still a matter of intensive debate.  The analysis of Pande et al. \cite{PandeCorrelations} argues in favor of a tendency for sequences to be annealed, i.e., to exhibit positive correlations in the hydrophilicity and anticorrelations in the charge of amino acids, which would suggest a bias towards the FRSB freezing scenario for proteins. However, the studies by Irb\"ack et al. \cite{Irbaeck96, Irbaeck97a, Irbaeck97b} rather point towards anticorrelations in the HP-type degrees of freedom which would favor a scenario with a direct transition from the liquid to the frozen glass. The discrepancies of these studies mainly concern the nature of long range correlations while on the level of nearest neighbor correlations, the protein sequences appear to be rather random, having $\pi \approx 1/2$ with respect to both charge and hydrophobic/hydrophilic degrees of freedom. It would be very interesting to understand whether the folding of natural proteins takes advantage from their sequences being very close to the critical border between the two scenarios. On the other hand, as mentioned earlier, most protein sequence design schemes tend to result in (partially) annealed monomer chains which are therefore
likely
to exhibit the intermediate soft glass phase.

%
%
\section{The close-packed limit}
\label{frozenphase}

In this section we provide a detailed analysis
of the frozen phase in the limit of high density.
We first show the existence of a special `REM-like' fully polarized 
solution of the 1RSB cavity equations at temperatures 
below the liquid's
entropy crisis.
Then we show that this solution is stable in the close-packed 
limit of high densities.
\subsection{A fully polarized solution}
\label{fullypolarized}
There always exists a `fully polarized' solution to the cavity equation (\ref{onestepcavity})
which describes pure states consisting of essentially one unique frozen polymer configuration. In each such state, a given site only admits one specific local conformation. On averaging over the different pure states, the given site will be found in conformation $\alpha$ with frequency $w_\alpha$. The local field distributions then take the form  
\be
	\label{polsolution}
	\rho_{\rm pol}({\bf p})=\sum_{\alpha}w_\alpha(\beta,m)\delta({\bf p}-{\bf e}^{(\alpha)}),
\ee

where the fields ${\bf e}^{(\alpha)}$ are defined by 
$e^{(\alpha)}_{\alpha'}=\delta_{\alpha\alpha'}$. This distribution solves 
the cavity equations when the frequencies $w_\alpha(\beta,m)$ coincide 
with the local fields of a liquid at the renormalized inverse temperature 
$\beta'=m\beta$, i.e., $w_\alpha(\beta,m)=p^*_\alpha(\beta'=m\beta)$.
The replicated free energy of this fully polarized solution is 
$\phi_1(\beta,m)=\phi_{\rm liq}(m\beta)$. 
The internal free energy of the corresponding frozen states is related to 
the liquid 
quantities via $f_{\rm pol}(\beta,m)=d(m\phi_1(\beta,m))/dm=u_{\rm liq}(m\beta)
-\mu\rho_{\rm liq}(m\beta)$, and the complexity of states is found from 
$\Sigma_{\rm pol}(\beta,m)=s_{\rm liq}(m\beta)$. As is evident from the 
nature of the pure states, their internal entropy vanishes. 

Let us for a moment postpone the discussion of the relevance of this solution,
and first discuss its physical interpretation. At each value of $\beta$ 
we  have to maximize $\phi_1$ over $0\le m\le 1$, under the condition 
$\Sigma\ge 0$. For temperatures above the liquid's entropy crisis, 
$\beta<\beta_{\rm cris}$, the maximum is attained at $m=1$ and we have 
$\omega_{g}=\omega_{\rm liq}$. When $\beta>\beta_{\rm cris}$, the static 
glass transition takes place and the free energy freezes to 
$\omega_g=\omega_{\rm liq}(\beta_{\rm cris})$, the Parisi parameter taking 
the value $m_s=\beta_{\rm cris}/\beta$. So this solution describes a full 
freezing of the polymer in 
some isolated specific configurations, taking place at 
$\beta=\beta_{\rm cris}$.
Notice that this scenario exactly parallels the one found in the REM.

Our numerical study of 
 the AB-copolymers in their highly frozen phase (beyond 
$\beta_d\approx 1.23$) finds a solution
$\rho({\bf p})$ which is close to the form (\ref{polsolution}), although small deviations persist, and the polarization is not complete. In the particular case of the alternating chain we numerically confirmed that the optimal Parisi parameter is well fitted by $m_s=T/T_s$ on the coexistence line.

\subsection{Stability analysis and the limit of maximal density} 
Up to this point we have not discussed the range of validity of the 
polarized solution (\ref{polsolution}), and in particular, its stability. 
Unfortunately, this is a difficult problem, and we only can provide 
partial answers.

The basic idea consists in perturbing the Ansatz (\ref{polsolution}) and 
checking whether the perturbation grows under the cavity iteration
(\ref{onestepcavity}). A simple perturbation consists in adding to 
(\ref{polsolution}) some `almost polarized' fields with a small total weight.
Namely we take a field distribution of the form
\begin{eqnarray}
\rho({\bf p}) = (1-a\vae)\rho_{\rm pol}({\bf p})+\vae\sum_{\alpha} 
w_\alpha \rho_{\alpha} ({\bf p})\, ,
\label{PerturbedPol}
\end{eqnarray}
where $\rho_\alpha({\bf p})$ is concentrated on fields ${\bf p}$ close
to ${\bf e}^{(\alpha)}$. In fact, it is more convenient to
think of it as a distribution over the `small' fields
$\vp \equiv \{ p_{\alpha'} \}_{\alpha'\neq \alpha}$. Hereafter,
we shall use the notation $\rho_{\alpha}(\vp)$ instead of 
$\rho_{\alpha}({\bf p})$. Finally notice that the $\rho_{\alpha}(\vp)$'s
need not to be normalized. Normalization is enforced by the constant $a$ 
in Eq. (\ref{PerturbedPol}).

Plugging the nsatz (\ref{PerturbedPol}) into Eq. (\ref{onestepcavity})
we get to linear order in $\epsilon$:
\begin{eqnarray}
\rho'_{\alpha_0}(\vp) = k\sum_{\alpha_1\dots\alpha_k} \!
P(\alpha_1\dots\alpha_k|\alpha_0)\int\! d\rho_{\alpha_1}(\vq)\;\;
\delta\left(\vp-\vI[\vq;\alpha_2\dots\alpha_k] \right)\, .\label{LinearizedPol}
\end{eqnarray}
Here we distinguished the distribution on the right-hand side, 
$\rho_{\alpha}(\cdot)$ from the one on the left-hand side
 $\rho_{\alpha}'(\cdot)$. In fact we are interested in the stability of
the iteration  (\ref{onestepcavity}) and not just in its fixed point.
Here $P(\alpha_1\dots\alpha_k|\alpha_0)$ is the probability
of finding conformations $\alpha_1\dots \alpha_k$ on the $k$ leaves of
the branch in Fig. \ref{figconformations}, 
constrained
to the root 
being in conformation $\alpha_0$. This must be computed
within the solution
described by Eq. (\ref{polsolution}) and can explicitly be 
written in terms of the weights $w_{\alpha}(\beta,m)$.
Finally $\vI[\vq;\alpha_2\dots\alpha_k]$ denotes the `small'
components of the cavity iteration:
 
$(\vI[\vq;\alpha_2\dots\alpha_k])_{\alpha} = I_{\alpha}
[{\bf q},{\bf e}^{(\alpha_2)}\dots{\bf e}^{(\alpha_k)}]$ for 
$\alpha\neq \alpha_0$.

Instead of continuing in full generality, let us consider the example of an
alternating F-model in the closed-packed limit 
with $E_{AA}=E_{BB}=-E_{AB}=-1$
(remember that in this case we found a 
discontinuous phase transition with a highly polarized low temperature phase,
cf. Sec. \ref{AlternatingHP}). Eqs. (\ref{LinearizedPol}) reduce
to
\begin{eqnarray}
\rho'_{1A}(\vp) & = & \sum_{n=0}^{k-1} f_n\,\delta(p_{2A},p_{2B})\,
\int\!\! d\rho_{1B}(\vq) \;\; \delta\left(p_{1B}-
 e^{-2\beta (k-1-2n)}q_{1A}\right) +\label{LinearizedAlt1}\\
&&+ g_A \delta(p_{1B},p_{2B})\,
\int\!\! d\rho_{2A}(\vq)\; \delta(p_{2A}- e^{-\beta}q_{1B})+
  g_B \delta(p_{1B},p_{2B})\,
\int\!\! d\rho_{2B}(\vq)\; \delta(p_{2A}- e^{\beta}q_{1B})\, ,\nonumber\\
\rho'_{2A}(\vp) & = & 2
\delta(p_{1B},p_{2B})\, \int\!\!d\rho_{1B}(\vq)\;\delta\left(
p_{1A}- e^{\beta}q_{2A}- e^{-\beta} q_{2B}\right)\, ,\label{LinearizedAlt2}
\end{eqnarray}
plus two equations obtained by interchanging $A$ and $B$.
Here we used the shorthand $\delta(x,y)= \delta(x)\delta(y)$ and
expanded $\vI[\vq;\alpha_2\dots\alpha_k]$ in  the delta functions 
to linear order in $q_{\alpha}$ for $\alpha\neq\alpha_1$.
The weights $\{ f_n\}$ and $g_{A/B}$ are given by
\begin{eqnarray}
f_n = \frac{1}{(2\cosh \beta m)^{k-1}}\, {k-1 \choose n}\, e^{-
\beta m(k-1-2n)}\, ,\;\;\;\;
g_{A/B} =\frac{k-1}{1+e^{\pm 2\beta m}}\, .
\end{eqnarray}

A little thought shows that, after one iteration of Eqs.
(\ref{LinearizedAlt1}), (\ref{LinearizedAlt2}) we can set
\begin{eqnarray}
\rho_{1A} (\vp) & = & \delta(p_{2A},p_{2B})\; \rho_{1A\to 1B}(p_{1B})
+ \delta(p_{1B},p_{2B})\; \rho_{1A\to 2A}(p_{2A})\, ,\\
\rho_{2A} (\vp) & = & \delta(p_{1B},p_{2B})\;\rho_{2A\to 1A}(p_{1A})\, .
\end{eqnarray}
and that the linearized recursions decouple in the three `sectors' 
$\{ 1A\to 1B, 1B\to 1A\}$, $\{ 1A\to 2A, 2B\to 1B\}$, 
$\{ 1B\to 2B, 2A\to 1A\}$. The first sector corresponds to shifts of the chain
and turns out to be marginally stable (the function
$\vI[\vq;\alpha_2\dots\alpha_k]$ has to be developed to second order in $\vq$).

The other two sectors correspond to structural rearrangements
of the backbone and become unstable when 
$m\beta<(m\beta)_c\equiv y_c=1/2\cdot\log(2k-3)$.  This instability  has 
 a simple physical interpretation. The pure states described by $(m\beta)_c$
have a free energy density $f_c=1/2$. This means that on average,  
a randomly chosen site has one violated neighboring bond, i.e., one 
neighbor occupied by a monomer of the opposite type. It is thus possible 
to rearrange the backbone of the alternating chain without paying energy 
by opening the chain at the given site and redirecting it in the direction 
of the violated bond, and propagating the rearrangement through the lattice,
see Fig.~\ref{Fig:rearrang}.

\begin{figure}
	\resizebox{9 cm}{!}{
  \includegraphics{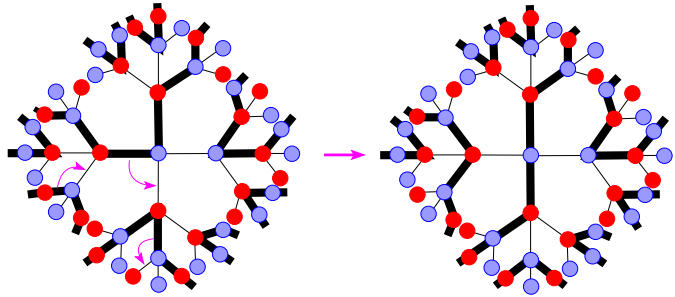}}
	\caption{Instability of the completely frozen solution due to rearrangements of the backbone.}
\label{Fig:rearrang}
\end{figure}

 For $k\leq 6$ the instability appears at a smaller value than the liquid 
entropy crisis, $y_c<\beta_{\rm cris}(\mu\ra \infty)$. Thus, at low 
temperatures, $\beta>y_c$, the thermodynamically relevant close packed 
states are correctly described by the stable polarized solution with 
$m_s=\beta_{\rm cris}/\beta$. In particular, we can immediately deduce 
the ground state energy
of Hamiltonian walks of an alternating HP-chain on a fixed connectivity 
random graph from the value  of 
$\phi_{\rm liq}(\beta_{\rm cris};\mu\ra \infty)$: This yields
 $0.083686, 0.120619, 0.172602, 0.236348$ violated bonds per site for 
$k=3,4,5$ and $6$, respectively.

A numerical study of the cavity recursion equations at maximal density
actually finds, for $k\leq 6$, a coexistence of the polarized solution with
another solution in some intermediate range $[y_c,y_t]$. 
This is a peculiarity of the infinite $\mu$ regime, 
the numerics at finite but large $\mu$ suggesting that the polarized solution is unphysical below $y_t$. However, since $y_t<\beta_{\rm cris}$, the polarized Ansatz still correctly describes the low temperature regime.

What happens away from the $\mu\to\infty$ limit? The possibility
of voids allows for new terms in the sum over conformations,
cf. Eq. (\ref{LinearizedPol}). It turns out that the iterations 
become unstable in the new sectors 
$\{0\to\ua a, \ua a \to 0\}$ and $\{0\to\da a, \da a \to 0\}$.  
Physically, this means that the presence 
of voids in the lattice always allows for a rearrangement of the polymer 
configuration in some (perhaps very rare) regions, preventing a complete 
freezing in a single state. Still, at $y\ge y_t$ a stable fixed point close to the polarized solution (\ref{polsolution}) exists. 

Let us finally notice that the stability of the polarized solution can be 
studied within a larger 2RSB Ansatz \cite{MontanariRicci03}. The results 
coincide with the simplified treatment presented here.
These results are further confirmed if one studies the behavior
of field distributions in the $T\to 0$ limit 
following Ref.~\cite{MontanariParisiRicci04}.

%
%
\subsection{Exact enumerations on a cube} 

In an attempt to verify the 1RSB or even REM-like nature of heteropolymers, 
Shakhnovich and Gutin have exactly enumerated all conformations of fully 
compact random 27-mers on a $3\times 3\times 3$ cube and calculated the 
overlap distribution function $P(q)$ as a function of temperature 
\cite{27mer,27mer93}. They interpreted their results in favor of a REM-like 
scenario where only a small number of states dominated the low temperature 
regime and $P(q)$ exhibited typical features of a discontinuous glass phase.
In view of our mean field predictions, one would expect to find a different 
scenario when repeating this analysis for copolymers (with a certain 
amount of sequence correlations) in their soft glass phase. 

We first repeated this enumeration study for \textit{random} ampholytes,
and found a $P(q)$ order parameter very similar to the 
random bond case studied originally \cite{27mer}, in 
agreement with the results of \cite{27mer93}. However the same
analysis done for correlated ampholytic sequences with various
values of $\pi$ did not show any clear dependence on $\pi$.
This absence of evidence can have two origins.
On the one hand it might be due to the extreme restrictions 
that full packing imposes on the conformations. 
We have seen above that the fully dense limit is very subtle 
since physically important degrees of freedom, which are found in a system with voids, 
are artificially suppressed, as has been put forward by many 
authors~\cite{nonergodicpolymer,Grosberg00, Timoshenko98}. 
On the other hand it seems that
these sizes are too small to study the true phase space 
structure of the glass phase. 

\section{Discussion and Conclusion}

The cavity method approaches the lattice heteropolymer problem from a new
point of view in that it analyzes the conformational degrees of freedom of
chains with quenched-in sequences. Furthermore, this method allows to study
the whole temperature range and describes the $\Theta-collapse$ and the low
temperature physics within the same formalism. In this sense we believe it 
provides an interesting new perspective in the analytic studies of heteropolymer
folding.

With this local approach we have studied the frustration effects on a 
given site of the lattice. We find that the  decisive features determining 
the nature of the low temperature physics are the short-range correlations 
in the monomer sequence. Polymers whose monomer distribution along the 
chain tends to be annealed have a proclivity
to undergo a continuous glass 
transition to a soft glass phase before the strong freezing transition 
takes place. In oppositely correlated sequences the freezing occurs 
directly from the liquid phase.
A weakly polarized phase with broken ergodicity and a high sensitivity 
to the specific sequence, has also been observed in the extensive numerical 
analysis of the phase diagram for specific hydrophilic/hydrophobic chains 
\cite{Timoshenko98}, and the qualitative differences found between selected 
sequences indeed reflect the general tendencies that we predict from the 
cavity analysis of the slightly different but closely related HP-like model.  

The temperature of the dynamic transition at which highly frozen pure 
states appear is almost independent of the correlations in the sequence as 
we found from the numerical solution of the 1RSB cavity equations. 
For the time being we do not have a deeper understanding of this finding, 
which is in accordance with numerical observations in the dynamics of 
copolymer folding. We hope to obtain better analytical insight into this 
phenomenon from a thorough analysis of the stability of the highly polarized 
low temperature states. This would probably also explain why the border 
between the 2RSB freezing scenario with an intermediate soft glass and 
the scenario of a direct transition liquid-frozen glass is so close to 
the Markov parameter $\pi=1/2$, corresponding to uncorrelated chains.

It would be interesting to verify the predictions for Markovian chains 
experimentally (preferably with ampholytes where the pair interactions 
are rather strong). In fact it is possible to 
fabricate Markovian copolymers from a random polymerization process, 
where $\pi$ can be controlled by changing the chemical parameters of the 
solution.
Furthermore, it will be very interesting to review the studies of sequence 
correlations in natural proteins in the light of our findings.  

The results of the cavity method are expected to be exact for polymers 
on random (Bethe) lattices, as is indeed corroborated by numerical 
simulations. However, on real space lattices the Bethe approximation 
neglects the correlations arising from small loops. 

It would thus be very important to check the effect of sequence correlations 
through numerical simulations of polymers on a cubic lattice,
using our mean field predictions as a guideline. 
One regime in which the small loops of the cubic lattice can 
yield a behavior which is qualitatively different from the present
mean field analysis is the case where
 the polymer has a strong tendency to form local crumples, 
as it happens in block copolymers which undergo a microphase separation. 
In order to study such problems analytically, it would be interesting to
improve the Bethe approximation by considering enlarged cavities that 
contain not only a single site but a small cluster of nearby sites. 
This actually amounts to a further step in the framework of the cluster 
variational method. For the homopolymeric case a first step 
in this direction has been carried out in \cite{Pretti02cactus}. 

Already on the level of the simplest copolymer model we found a 
surprisingly rich phase diagram as a function of temperature and 
sequence correlations. But clearly, the cavity method is amenable to a 
number of generalizations that allow to study more sophisticated models 
of biopolymers, including for instance backbone stiffness, orientational 
degrees of freedom, or additional structural constraints such as the 
saturation of monomer-monomer interactions, which are crucial, e.g., 
for the folding of RNA. 
%
%

\appendix

\section{Finding the liquid solution}
\label{LiquidApp}

In this Appendix we show how the translation invariant liquid solution can
be found by solving a set of $|{\cal A}|+2$ equations (instead of 
$3L+1$ equations as it may appear from Eqs. (\ref{cavity0})-(\ref{cavity2})).
First of all it is convenient to make a change of variables defining
\begin{eqnarray}
w^{(i)}_0 \equiv \sum_{a=1}^L \,\frac{p_{2a}^{(i)}}{p_0^{(i)}}\, ,
\;\;\;\;\;\;
z^{(i)}_{\uparrow a} \equiv \frac{p^{(i)}_{\uparrow a}}{p^{(i)}_{0}}\, ,
\;\;\;\;\;\; 
z^{(i)}_{\downarrow a} \equiv \frac{p^{(i)}_{\downarrow a}}{p^{(i)}_{0}}
\;\;\;\;\;\;
w^{(i)}_{\sigma} \equiv \sum_{a=1}^L\,
\frac{p_{2a}^{(i)}}{p_0^{(i)}}\exp(-\beta E_{\sigma,\sigma_a})\, .
\end{eqnarray}
It is easy to see that the cavity equations
(\ref{cavity0})-(\ref{cavity2})), the free energy (\ref{FreeEnergy}) and all
the others observables, can be rewritten in terms of these 
$2L+|{\cal A}|+1$ variables. In using the new variables, when not
specified, we shall assume that the index 
$\sigma$ belongs to the enlarged space $\{ 0\}\cup {\cal A}$. 
We will set $E_{0,\sigma}=E_{\sigma,0}=0$.

The liquid fixed point has the translation invariant form 
$w^{(i)}_{\sigma} = w_{\sigma}$,  $z^{(i)}_{\uparrow a}= z_{\uparrow a}$,
$z^{(i)}_{\downarrow a}= z_{\downarrow a}$. The corresponding equations
are easily written:
\begin{eqnarray}
z_{\uparrow a} & = & k e^{\beta\mu}\frac{z_{\uparrow,a+1}}{1+w_0}\,
\left(\frac{1+w_{\sigma_a}}{1+w_0}\right)^{k-1}\, ,\label{liquid1}\\
z_{\downarrow a} & = & k e^{\beta\mu}\frac{z_{\downarrow,a-1}}{1+w_0}\,
\left(\frac{1+w_{\sigma_a}}{1+w_0}\right)^{k-1}\, ,\label{liquid2}\\
w_{\sigma} & = & k(k-1)e^{\beta\mu} \sum_{a=1}^L \, 
e^{-\beta E_{\sigma,\sigma_a}}\frac{z_{\uparrow,a+1}}{1+w_0}
\frac{z_{\downarrow,a-1}}{1+w_0}\left(\frac{1+w_{\sigma_a}}{1+w_0}\right)^{k-2}\, .
\label{liquid3}
\end{eqnarray}
It is important to notice that the above equations are invariant 
under the transformation $z_{\uparrow a}\to \gamma\cdot z_{\uparrow a}$,
$z_{\downarrow a}\to \gamma^{-1}\cdot z_{\downarrow a}$ for any
positive $\gamma$: we shall fix this freedom below.
The reader can easily check that any physical 
observable (such as the free energy, the local energy or the local 
density) is also invariant under such a transformation. This happens because,
when following the chain along its conventional direction, each time 
we arrive at a site $i$, we are obliged to leave the site as well.

The above equations admit of course the trivial coil solution 
$z_{\uparrow a} = z_{\downarrow a} = 0$. Moreover, if one has
$z_{\uparrow a_0}= 0$ ($z_{\downarrow a_0}= 0$) for a particular $a_0$,
this implies $z_{\uparrow a}= 0$ ($z_{\downarrow a}= 0$) for any $a$.
Therefore, we shall hereafter assume that $z_{\uparrow a}, z_{\downarrow a}
\neq 0$ for any $a$. In this case Eqs. (\ref{liquid1})-(\ref{liquid2})
imply the consistency condition
\begin{eqnarray}
1 = \left(\frac{k e^{\beta \mu}}{1+w_0}\right)^L\prod_{\sigma\in{\cal A}}
\left(\frac{1+w_{\sigma}}{1+w_0}\right)^{(k-1)L\nu_{\sigma}}
\label{consistency}
\end{eqnarray}

Equations (\ref{liquid1}) and (\ref{liquid2}) are easily solved:
\begin{eqnarray}
z_{\uparrow a} & =& \prod_{b=a}^{L} \frac{k e^{\beta\mu}}{1+w_0}
\left(\frac{1+w_{\sigma_b}}{1+w_0}\right)^{k-1}\, z_{\uparrow}\, ,
\label{up_sol}\\
z_{\downarrow a} & =& \prod_{b=1}^{a} \frac{k e^{\beta\mu}}{1+w_0}
\left(\frac{1+w_{\sigma_b}}{1+w_0}\right)^{k-1}\, z_{\downarrow}\, ,
\label{down_sol}
\end{eqnarray}
where $z_{\uparrow}$, $z_{\downarrow}$ are two integration constants.
We can  exploit the invariance mentioned above in order to fix
$z_{\uparrow} = z_{\downarrow} = z$.

Plugging the expressions  (\ref{up_sol}), (\ref{down_sol}) into 
Eq. (\ref{liquid3}), and using Eq. (\ref{consistency}), we get
\begin{eqnarray}
w_{\sigma} = (k-1)  L z^2\, \sum_{\tau\in {\cal A}}
\frac{\nu_{\tau} e^{-\beta E_{\sigma\tau}}}{1+w_{\tau}}\, .
\label{Aplus1}
\end{eqnarray}
We are therefore left with a set of $|{\cal A}|+2$ equations 
(Eq. (\ref{consistency}) plus the $|{\cal A}|+1$  equations 
in (\ref{Aplus1})) for $|{\cal A}|+2$ real variables ($z$ and the 
 $|{\cal A}|+1$  variables $w_{\sigma}$). As anticipated
these equations depend on the sequence just through the frequencies
$\nu_{\sigma}$, $\sigma\in{\cal A}$. The reader will easily check that the
same is true for any physical observable.

Near the $\Theta$ point all $w_{\sigma}$ are small, and (\ref{Aplus1}) shows that to lowest order they satisfy $w_{\sigma}\approx w_0 \sum_{\tau\in {\cal A}}\nu_\tau  e^{-\beta E_{\sigma\tau}}$. By imposing that a non-trivial solution of (\ref{consistency}) should exist one immediately obtains the equation (\ref{thetapoint}) for the location of the $\Theta$ point.
%
%
\section{Neutral $AB$-copolymers: local stability analysis}
\label{Appendixinstab}

Here we outline the computation of the local stability condition
for an $AB$ copolymer having a generic period-$L$ sequence. 
We shall use, depending on the context, the two notations
$\sigma_a\in\{A,B\}$, or $\sigma_a\in\{+,-\}$ for the sequence.

As already mentioned in Sec. \ref{LocalSection}, we consider the case of 
an interaction matrix symmetric under $A\leftrightarrow B$ interchange.
Without loss of generality, we can restrict ourselves to 
the cases of the AF- and F-models defined in Sec. \ref{DefinitionSection}.
Moreover we shall assume that the sequence is neutral, i.e.
$\nu_A = \nu_B = 1/2$. Under these hypothesis, 
Eqs. (\ref{liquid1})-(\ref{liquid3}) admit the symmetric solution 
$z_{\uparrow a} = z_{\downarrow a} = z/\sqrt{L}$, $w_0 = w$,
$w_a= w\cosh \beta$, where $z$ and $w$ are determined by solving
the equations
\begin{eqnarray}
z & = & k e^{\beta\mu}\left(\frac{z}{1+w}\right)\left(
\frac{1+w\cosh\beta}{1+w}\right)^{k-1}\, ,\label{liquidAB1}\\
w & = & k(k-1)e^{\beta\mu}\,\left(\frac{z}{1+w}\right)^2
\left(\frac{1+w\cosh\beta}{1+w}\right)^{k-2}\, .\label{liquidAB2}
\end{eqnarray}

We want to compute the local stability of the cavity recursions 
(\ref{cavity0})-(\ref{cavity2}) around the above solution. 
We therefore imagine that the cavity fields for one
of the sites $1, \dots k$ (let us say the site $1$) have been 
slightly perturbed and compute the effect of such a perturbation on 
the site 0. To linear order we get:
\begin{eqnarray}
\delta z^{(0)}_{\uparrow a} &=& A\, \delta z^{(1)}_{\uparrow, a+1} -
B\,\delta w^{(1)}_0 +C\, \delta w^{(1)}_{\sigma_a}\, ,
\label{linear1}\\
\delta z^{(0)}_{\downarrow a} &=& A\, \delta z^{(1)}_{\downarrow, a-1} -
B\, \delta w^{(1)}_0 +C\, \delta w^{(1)}_{\sigma_a}\, ,\label{linear2}\\
\delta w^{(0)}_0 & = & D\sum_{a=1}^L(\delta z^{(1)}_{\uparrow a}+
\delta z^{(1)}_{\downarrow a})-E\, \delta w^{(1)}_0 + F\, 
\sum_{\sigma\in\{A,B\}}\delta w^{(1)}_{\sigma}\, ,\label{linear3}\\
 \delta w^{(0)}_{\sigma} & = & G\sum_{a=1}^L(e^{-\beta E_{\sigma \sigma(a-1)}}
\delta z^{(1)}_{\uparrow,a} + e^{-\beta E_{\sigma \sigma(a+1)}}
\delta z^{(1)}_{\downarrow,a}) - H\, \delta w^{(1)}_0+F\, \sum_{\tau \in \{A,B\}}
e^{-\beta E_{\sigma\tau}}\,\delta w^{(1)}_{\tau}\, .\label{linear4}
\end{eqnarray}
The constants $A$--$H$
are all positive, and can be expressed in terms 
of the solution of Eqs. (\ref{liquidAB1})-(\ref{liquidAB2}). In the following
we will just need  the combinations below:
\begin{eqnarray}
A = \frac{1}{k}\, ,\;\;\;\;\;\;\;\;
CG = \frac{k-1}{k^2L}\frac{w}{1+cw}\, ,\;\;\;\;\;\;\;\;
F = \frac{k-2}{2k}\frac{w}{1+cw}\, ,
\end{eqnarray}
where we used the shorthand $c\equiv\cosh\beta$.

We must now identify the most relevant perturbation, i.e., the largest 
eigenvalue of the linear transformation (\ref{linear1})-(\ref{linear4}).
It is simple to show that the subspace 
\begin{eqnarray}
\label{subspace}
\left\{ \delta w_0=0;\; \sum_{a=1}^L \delta z_{\uparrow a} = 0;\;
\sum_{a=1}^L \delta z_{\downarrow a} = 0;\; \sum_{\sigma\in \{ A, B\}}\;
\delta w_{\sigma} = 0\right\}
\end{eqnarray}
is preserved by the iteration (\ref{linear1})-(\ref{linear4}).
It can be shown that the most relevant eigenvector lies indeed within 
this subspace. We restrict to it by defining the variables
\begin{eqnarray}
\delta\overline{w} \equiv \delta w_A -\delta w_B\, ,\;\;\;\;\;
\delta_{\uparrow b} \equiv \sum_{a=1}^L\delta z_{\uparrow a}\, \sigma_{a-b}\,
\;\;\;\;
\delta_{\downarrow b} \equiv \sum_{a=1}^L\delta z_{\downarrow a}\, 
\sigma_{a+b}\, ,
\end{eqnarray}
where we used $\sigma_a\in\{+,-\}$ for the polymer sequence.
Using the new variables we can rewrite the iteration 
(\ref{linear1})-(\ref{linear4}) as follows:
\begin{eqnarray}
\delta^{(0)}_{\uparrow a} & = & A\,  \delta^{(1)}_{\uparrow, a+1}+
\frac{L}{2}C q_{-a}\, \delta\overline{w}^{(1)}\, ,
\label{liquidABred1}\\
\delta^{(0)}_{\downarrow a} & = & A\,  \delta^{(1)}_{\downarrow, a+1}+
\frac{L}{2}C q_{a}\, \delta\overline{w}^{(1)}\, ,\label{liquidABred2}\\
\delta\overline{w}^{(0)} & = & 
2Gs\, (\delta^{(1)}_{\uparrow 1}+\delta^{(1)}_{\downarrow 1})
+2Fs\, \delta\overline{w}^{(1)}\, ,\label{liquidABred3}
\end{eqnarray}
where we introduced the notation $s\equiv \sinh\beta$ (for
the F-model) or $s \equiv -\sinh\beta$ (for the AF-model), and the sequence
correlation function
\begin{eqnarray}
q_b = \frac{1}{L}\sum_{a=1}^L \, \sigma_a\sigma_{a+b}\, .
\end{eqnarray}

Notice that $q_b=q_{-b}$. This remark allows us to sum 
Eqs.~(\ref{liquidABred1}) and  (\ref{liquidABred2}) and to introduce
the Fourier transform (for $p=2\pi n/L$, $n\in\{1,\dots, L-1\}$):
\begin{eqnarray}
\delta(p) = \sum_{a=1}^{L}\, (\delta_{\uparrow,a}+\delta_{\downarrow,a})
\, e^{-ipa}\, .
\end{eqnarray}
We obtain therefore
\begin{eqnarray}
\delta^{(0)}(p) & = & Ae^{ip}\,  \delta^{(1)}(p)+
LC q(p)\, \delta\overline{w}^{(1)}\, ,\\
\delta\overline{w}^{(0)} & = & 
2Gs\, \frac{1}{L}\sum_{p}\delta^{(1)}(p)\,e ^{ip}
+2Fs\, \delta\overline{w}^{(1)}\, .
\end{eqnarray}
We can now set $\delta^{(0)}(p) = \lambda \delta^{(1)}(p)$, 
$\delta\overline{w}^{(0)} = \lambda \delta\overline{w}^{(1)}$, and solve
for $\lambda$, thus recovering Eq. (\ref{ABinstab}).
%
%
\section{Coexistence condition for a many states molecule}
\label{ap:coex}
It may be
interesting to explicitly treat the case of an isolated molecule in 
equilibrium with the solvent and determine the coexistence condition in the glass phase. The result is not obvious since the 
system can exist in many different states $\gamma\in\{ 1,\dots {\cal N}\}$
with (extensive) grand potential $\{\Omega_{1},\dots,\Omega_{\cal N}\}$.
Each one of these states describes a molecule confined to a volume $V$.

Let us suppose that each state can be traced as the volume $V$ of
the system is changed. This gives us the volume-dependent potentials
$\Omega_{\gamma}(V)$. If the state $\gamma$ is to describe a 
molecule in equilibrium with the solvent it should exert no 
pressure on the walls of the container:
\begin{eqnarray}
\frac{d\Omega_{\gamma}}{d V}=0\, .\label{ZeroPressure}
\end{eqnarray}

We want to compute the typical value of the above quantity for states 
having a certain free-energy density: $\Omega_{\gamma}\approx V\omega$.
Let us step back for a moment and 
consider the extensive complexity $\widehat{\Sigma}(\Omega;V,\mu)$, where
we made explicit the dependence upon the volume $V$ and the chemical potential
$\mu$. If 
we assume that states do not bifurcate and do not die
(or come into existence) as the volume is changed, it is easy to show that
\cite{LopatinIoffe02}, for almost any state $\gamma$:
\begin{eqnarray}
\label{equalstates}
\widehat{\Sigma}(\Omega_{\gamma}+d\Omega_{\gamma};V+dV,\mu) = 
\widehat{\Sigma}(\Omega_{\gamma};V,\mu)\, .
\end{eqnarray}
Using the asymptotic behavior 
$\widehat{\Sigma}(\Omega;V,\mu)\approx V\Sigma(\omega,\mu)$, 
and the general relations
from Sec.~\ref{sect:mdef}
 we can establish the coexistence condition 
either in the $(m,\mu)$ or in the $(\omega,\mu)$ plane (we always
assume $\beta$ and the energy parameters to be fixed).
From (\ref{equalstates}), we immediately obtain the condition in the $(\mu,\omega)$ plane:
\begin{eqnarray}
\omega\frac{\partial\Sigma}{\partial\omega}(\omega,\mu) = 
\Sigma(\omega,\mu)\, .\label{Coexistence}
\end{eqnarray}
This is suggestive of a balance between an ``internal'' osmotic pressure,
 $\omega$, and an ``interstate'' pressure ($\Sigma/\partial_{\omega}\Sigma$).
 In the $(m,\mu)$ plane, the condition assumes a more compact form
$\phi_1(m,\mu)=0$. 
If we consider the lowest lying states, their free energy density
$\omega_{\rm s}(\mu)$ is determined by the vanishing of the complexity:
$\Sigma(\omega_{\rm s}(\mu),\mu) = 0$. Therefore Eq.~(\ref{Coexistence}) 
is satisfied for $\mu = \mu_{\rm s}$, with $\omega_{\rm s}(\mu_{\rm s})=0$.
This coincides with the condition for a unique pure state. 
If metastable states are considered,
Eq.~(\ref{Coexistence}) receives a non-vanishing contribution from the complexity:
in particular, one obtains $\omega>0$. This is quite striking since we 
did not assume the system to equilibrate among states of a given free-energy
(which indeed does not happen on the short time scales that are relevant 
to determine the boundary conditions with the solvent).

\begin{figure}
\centerline{
\epsfig{figure=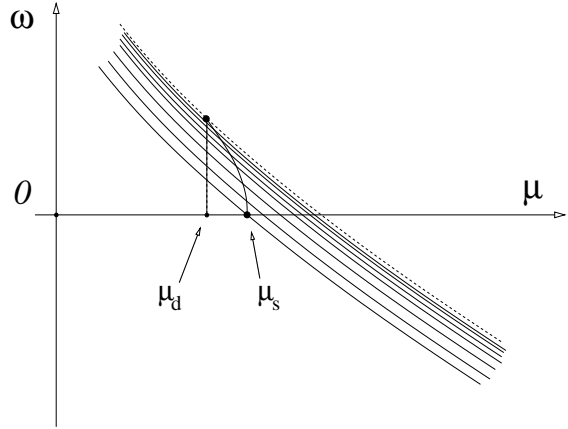,width=0.45\linewidth}}
\caption{A schematic view of the coexistence between a multi-state molecule
and the solvent. Each line represents the evolution of the internal
free energy of a state as the volume is changed 
($\Sigma(\omega,\mu)=const.$).
The thick line shows the states which are in equilibrium with the solvent. In particular,
we signal the coexistence chemical potentials for static and dynamic states.}
\label{CoexScheme}
\end{figure}
In Fig.~\ref{CoexScheme} we represent the condition (\ref{Coexistence}) 
in the $(\omega,\mu)$ plane. Notice that in general metastable states (with
$\Sigma>0$) on the coexistence line correspond to lower chemical potential 
than that of thermodynamically relevant states.

Let us finally consider the coexistence line at thermodynamic equilibrium. 
Dominant states are obtained by minimizing the free energy 
$\omega-\beta^{-1}\Sigma(\omega,\mu)$ with respect to $\omega$. The 
coexistence chemical potential $\mu_*$ is then obtained from 
Eq.~(\ref{Coexistence}). In a more compact (but formal) way, it is 
determined from the condition
\begin{eqnarray}
\max \{\phi_1(m,\mu_*)|m\in[0,1]\} = 0\, .\label{EquilibriumLine}
\end{eqnarray}
In the main body of the paper we focus on the behavior of the polymer 
on this line. Generally speaking, at high temperature the maximum in 
Eq.~(\ref{EquilibriumLine}) is attained at $m=1$. Since 
$\phi_1(m=1,\mu) = \phi_{\rm liq}(\mu)$, in this region the coexistence 
line is the same as for the liquid phase. At lower temperatures the 
maximum is attained for $0<m_*<1$ and the thermodynamic coexistence line 
lies above the liquid one. We refer to Fig.~\ref{phasedia} for a summary of this behavior.

%
%
\section{Expansion of moments at the continuous glass transition}
\label{Appendixmoments}

Here we analyze the solution of the cavity recursion near the continuous transition to first non-trivial order in an expansion of its moments.

Using both sides of the cavity recursion equation on the 1RSB level 
(\ref{onestepcavity}) in order to calculate the moments of the cavity 
fields, one obtains a set of coupled non-linear equations for the moments 
of the fields $p_{\alpha}$ over the distribution $\rho({\bf p})$. 
It is convenient to change coordinates and define the fields
$\Delta_{\mu} = \sum_{\alpha} A_{\mu}^{\, \alpha}
(p_{\alpha}-p^*_{\alpha})$ in such a way as to diagonalize
the matrix (\ref{Mdef}). Hereafter we shall denote by $\mu=1$
the most instable (`replicon') direction in this matrix, and by $\lambda$ 
the corresponding eigenvalue. 

A careful 
analysis allows to establish the scaling of the moments with respect to the 
small parameter $k\lambda^2-1\sim T_i-T$ close to the instability 
(\ref{instability}). The leading moment is given by the second moment 
of the replicon mode. One finds $\<\Delta_1^2\>\sim (T_i-T)$ (the brackets 
$\<\>$ denote the average with respect to $\rho$), while all other moments 
of deviations from the liquid fixed point ${\bf p}^*$ are at least of second 
order in $T_i-T$. The only moments of order $(T_i-T)^2$ are the first 
order moments of $\<\Delta_{\mu}\>$, the remaining second moments 
$\<\Delta_1\Delta_a\>$, $\<\Delta_{\mu}\Delta_{\nu}\>$, with 
$\mu, \nu>1$  and the higher moments $\<\Delta_1^3\>$,
$\<\Delta_1^2\Delta_\mu \>$ and $\<\Delta_1^4\>$.  

We will exploit the knowledge about this scaling in order to expand 
the 1RSB free energy (\ref{onestepphi}) in powers of $T_i-T$ around 
the liquid solution. 

The site term gives rise to a series 
\bea
        \nonumber
	\log \left[\left(w_s^{\rm liq} \right)^m \left( 1 +
	\left(\begin{array}{c} 	m\\1 \end{array}\right) \< \sum_{i=1}^{k+1} w_{s,\mu}\Delta^{(i)}_{\mu}+\frac{1}{2}\sum_{i\neq j}^{k+1}w_{s,\mu\nu}\Delta^{(i)}_{\mu}\Delta^{(j)}_{\nu} + \frac{1}{6}\sum_{i\neq j\neq l}^{k+1}w_{s,\mu\nu\rho}\Delta^{(i)}_{\mu}\Delta^{(j)}_{\nu}\Delta^{(l)}_{\rho}\dots \> \right. \right.\\
		\left.\left. +\left(\begin{array}{c} 	m\\2 \end{array}\right) \< \left(\sum_{i=1}^{k+1}w_{s,\mu}\Delta^{(i)}_{\mu}+\frac{1}{2}\sum_{i\neq j}^{k+1}w_{s,\mu\nu}\Delta^{(i)}_{\mu}\Delta^{(j)}_{\nu} +\frac{1}{6}\sum_{i\neq j\neq l}^{k+1}w_{s,\mu\nu\rho}\Delta^{(i)}_{\mu}\Delta^{(j)}_{\nu}\Delta^{(l)}_{\rho}\dots \right)^2\>+ \dots \right) \right]\nonumber\,,\\
\label{expansion}
\eea

where summation over direction indices $\mu,\nu,\rho=1,\dots,3L$ is tacitly understood and we used the shorthand notation 
\bea
	w_{s,\mu}=\left[\frac{1}{w_s}\frac{\partial w_s}{\partial\Delta^{(1)}_\mu}\right]_{\rm liq}, \\
	w_{s,\mu\nu}=\left[\frac{1}{w_s}\frac{\partial^2 w_s}{\partial\Delta^{(1)}_\mu\partial\Delta^{(2)}_\nu}\right]_{\rm liq}, \\
	w_{s,\mu\nu\rho}=\left[\frac{1}{w_s}\frac{\partial^3 w_s}{\partial\Delta^{(1)}_\mu\partial\Delta^{(2)}_\nu\partial\Delta^{(3)}_\rho}\right]_{\rm liq}\,.
\eea
Note that we have made use of the fact that $w_s$ is a 
multi-linear function of the fields ${\bf p}^{(i)}$ so that higher derivatives have to occur with respect to variables on different sites. The average $\<\>$ is with respect to the distributions $\rho({\bf p}^{(1)}),\dots,\rho({\bf p}^{(k+1)})$ on all sites.
The link term has an analogous expansion as (\ref{expansion}), but the triple sum vanishes since there are only two different field variables.
 
To proceed, we note the identity
\be
	\label{identity}
	w_s({\bf p}^{(1)},\dots,{\bf p}^{(k+1)})=C[{\bf p}^{(2)},\dots,{\bf p}^{(k+1)}]w_l({\bf p}^{(1)},I[{\bf p}^{(2)},\dots,{\bf p}^{(k+1)}]),
\ee
from which one immediately deduces 
\be
\label{firstder}
w_{s,\mu}=w_{l,\mu}
\ee
for all directions $\mu$. Using that $\partial C/\partial \Delta^{(2)}_1|_{\rm liq}=0$, as follows from the properties of the subspace (\ref{subspace}) which the replicon belongs to, one further finds 
\be
\label{w1=0}
w_{s,1}= w_{l,1}=0 
\ee
and 
\be
\label{w2}
w_{s,1\nu}=\lambda w_{l,1\nu}\,.
\ee 

Let us now discuss the terms that appear to increasing order in $T_i-T$ 
in the expansion of the free energy. There is no first order term 
proportional to $\<\Delta_1^2\>$, because of (\ref{w1=0}).
The second order terms $\<\Delta_\mu\>$, $\<\Delta_\mu\Delta_\nu\>$, 
$\<\Delta_1^3\>$, and $\<\Delta_1^2\Delta_\mu\>$ come with products 
of factors $w_{s/l,\mu}$ and cancel exactly between the site and link 
contributions, due to (\ref{firstder}). The only remaining term to 
second order is
\be
	\label{2order}
	\beta m\, \phi^{(2)}= \left(\begin{array}{c} m\\2 \end{array}\right)
	\left[-\frac{(k+1)k}{2}\left(w_{s,11}\right)^2 +\frac{k+1}{2}\left(w_{l,11}\right)^2\right] \<\Delta_1^2\>^2.
\ee
However, using (\ref{w2}), the coefficient in brackets is seen to be of order $k\lambda^2-1\sim T_i-T$.

The same mechanism suppresses the a priori third order terms 
$\<\Delta_1^2\>\<\Delta_\mu\Delta_\nu\>$ and 
$\<\Delta_1^2\>\<\Delta_{1}^2\Delta_\nu\>$ by an additional factor 
$k\lambda^2-1$ while the terms $\<\Delta_1^2\>\<\Delta_\mu\>$, 
$\<\Delta_1^3\>\<\Delta_1^2\>$ and $\<\Delta_1^4\>\<\Delta_1^2\>$ do not 
appear, again because of (\ref{w1=0}). The only surviving third order 
contributions are the site terms proportional to $\<\Delta_1^2\>^3$,
\be
	\label{3order}
	\beta m\, \phi^{(3)}= -(k+1)k(k-1)
	\left[\frac{1}{6}{m \choose 2} w_{s,111}^2
+{m \choose 3}w_{s,11}^3\right]\<\Delta_1^2\>^3.
\ee

To first non trivial order we finally have
\bea
	\label{phivarfinal}
	\phi_1&=& \phi_{\rm liq}+\phi^{(2)}+\phi^{(3)}+O((k\lambda^2-1)^4)\\
&=&\phi_{\rm liq}+\frac{1-m}{4\beta}(k+1)k\,
w_{s,11}^2({k\lambda^2}-1)\<\Delta_1^2\>^2 \nonumber \\
	&& +\frac{1-m}{2\beta}(k+1)k(k-1)
	\left(\frac{1}{6}w_{s,111}^2 -\frac{2-m}{3}w_{s,11}^3
\right)\<\Delta_1^2\>^3+O((k\lambda^2-1)^4).\nonumber
\eea

Since the form (\ref{onestepphi}) of the free energy is variational, we may obtain the leading moment to first order in $T_i-T$ from the condition $\partial \phi_1/\partial\<\Delta_1^2\> =0$,
\be 
	\label{D_1^2}
	\<\Delta_1^2\>=\frac{1}{(k-1)}\,\frac{w_{s,11}^2}{\left[(2-m)w_{s,11}^3 -w_{s,111}^2/2 \right]}(k\lambda^2-1)+O\left((k\lambda^2-1)^2\right)\,.
\ee
Plugging this result in Eq. (\ref{phivarfinal}) we get
\bea
\phi_1 &= & \phi_{\rm liq}+ \frac{(1-m)}{12\beta}\, \frac{k(k+1)}{(k-1)^2}\,
\frac{w_{s,11}^6}{\left[(2-m)w_{s,11}^3 -w_{s,111}^2/2 \right]^2}\,
(k\lambda^2-1)^3 +O((k\lambda^2-1)^4)\, .
\eea

Note that the prefactor of $(k\lambda^2-1)$ in Eq. (\ref{D_1^2})
has to be positive for consistency. 
A negative value indicates that there is no stable solution
close to the liquid fixed point and the glass transition would be
discontinuous. 
By explicit calculation of this coefficient at the instability point
we found this to happen only in very atypical sequences with
highly non-symmetric interactions. 

Evaluating the coefficients $w_{s,11}$ and $w_{s,111}$ requires the 
knowledge of the replicon eigenvector. This can be derived 
for the case of copolymers with symmetric interaction matrix
$E_{AA}= E_{BB} = -E_{AB}$, and equally frequent monomer species,
$\nu_A=\nu_B=1/2$, extending the arguments of  App.~\ref{Appendixinstab}.
In particular we obtain (using the variables defined in
Apps.~\ref{LiquidApp} and \ref{Appendixinstab}):
\begin{eqnarray}
\delta w_0 = 0\, ,\;\;\; \delta z_{\ua a} = \frac{C \delta w}{2\lambda}\sum_{n=0}^{\infty} 
\sigma_{a+n} (k\lambda)^{-n}\, , 
\;\;\; \delta z_{\da a} = \frac{C \delta w}{2\lambda}\sum_{n=0}^{\infty} 
\sigma_{a-n} (k\lambda)^{-n}\, ,\;\;\;
\delta w_{\sigma} = \sigma\delta w\, .\label{RepliconEvector}
\end{eqnarray}

Things simplify considerably in several important cases: $(i)$ alternating
copolymers; $(ii)$ anti-palindromic sequences; $(iii)$ Markov sequences in 
the $L\to\infty$ limit. In all this cases the ratio $w_{s,111}^2/w_{s,11}^3$
vanishes. The basic reason is that, because of Eq. (\ref{RepliconEvector}), 
$w_{s,111}$ turns out to be an odd function of  $\{\sigma_a\}$.
In these cases the free energy $\phi_1(m)$ takes the simpler form,
cf. (\ref{1orderphi}),

\begin{equation}
 \phi_1-\phi_{\rm liq}= \frac{(k+1)k^2}{12(k-1)^2 \beta} \frac{1-m}{(2-m)^2} (k\lambda^2-1)^3+ O\left((k\lambda^2-1)^4\right).
\end{equation} 
At the glass transition the maximum of $\phi_1$ is attained at 
$m_s=0$. The fourth order term will shift its position to $m_s\propto
k\lambda^2-1\sim T_i-T$, as we have explicitly checked in the alternating AB-ampholyte.
%
%
\section{Computing the order parameter in the cavity method} 
\label{app:op}
We show here how to compute the local structural order parameters 
(\ref{orpar_space}) using the cavity method.

In the spirit of the Bethe-Peierls approximation we treat the 
self-avoidance of the polymer chain just on a local level, forbidding 
it to leave a site on the edge on which it arrived, but neglecting 
further constraints that arise on a real space lattice. In the following, 
we call ``non reversal random walks'' (NRRW) this restricted class of 
walks on the cubic  lattice.

Let us rewrite the distance vector between monomers $i$ and $i+d$ 
as $\vec{R}^{(1)}_{i+d}-\vec{R}^{(1)}_{i}=\sum_{n=1}^{d}\vec{r}^{(1)}_n$ 
with $ \vec{r}^{(1)}_n=\vec{R}^{(1)}_{i+n}-\vec{R}^{(1)}_{i+n-1}$. 
If the positions along the chain are statistically equivalent,
the overlap $\<F_d\>_{\rm state}$ can be written as
\be 
	\label{Faverage}
	\<F_d^{(1,2)}\>_{\rm state}=\<\sum_{l=0}^{d}\left(\sum_{n=1}^{l} \vec{r}_n+\sum_{n=l+1}^{d} \vec{r}^{(1)}_n\right)\cdot
	\left(\sum_{n=1}^{l} \vec{r}_n+\sum_{n=l+1}^{d} \vec{r}^{(2)}_n\right)\>_{\rm state}
\ee
where we split the sum according to the length $l$ over which the replicas 
stay together and put $\vec{r}_n^{(1)}=\vec{r}_n^{(2)}=\vec{r}_n$ for 
$n\le l$. 
Note that once $l$ is fixed the common part of the path and the two 
legs of length $d-l$ can be considered as non reversal random walks, 
only subject to the constraint that the legs leave in different 
directions at the bifurcation. 
These random walks have all the same weight when averaging over 
pure states. Hence, in order to evaluate (\ref{Faverage}) it is 
sufficient to calculate  the probability $\<P(l)\>_{\rm state}$ for 
two replicas in the same state to follow the same path over a distance $l$, 
from which we obtain 
\be  
	\label{Fav2}
	\<F_d^{(1,2)}\>_{\rm state}=\sum_{l=0}^{d} \<P(l)\>_{\rm state} f(l;d)
\ee
where 
\be 
	f(l;d) = \<\left(\sum_{n=1}^{l} \vec{r}_n+\sum_{n=l+1}^{d} \vec{r}^{(1)}_n\right)\cdot
	\left(\sum_{n=1}^{l} \vec{r}_n+\sum_{n=l+1}^{d} \vec{r}^{(2)}_n\right)\>_{NRRW(l)}\,,
\ee
the average being taken over the uniform distribution of two NRRW's
after $l$ common links. 

Using that in a NRRW one has $\<\vec{r}_{n_1}\cdot \vec{r}_{n_2}\>_{NRRW}=
1/k^{|n_1-n_2|}$, and distinguishing the different possible 
conformations at the bifurcation, one easily finds
\bea 
	\label{f(l)}
	f(l;d)=l+2\sum_{j=1}^{l-1} \frac{l-j}{k^j} +\frac{2}{k}\sum_{n_1=1}^{l}\sum_{n_2=1}^{d-l}\frac{1}{k^{n_1+n_2-2}} -\frac{1}{k} \sum_{n_1=1}^{d-l}\sum_{n_2=1}^{d-l}\frac{1}{k^{n_1+n_2-2}}.
\eea
(The first two terms stem from the self overlap of the common part, 
the term in the middle is the cross term between the common part and 
a leg that continues straight with respect to the common part, and the 
last term is a negative contribution due to two legs leaving in opposite 
directions.)

In the liquid state, $\<P(l)\>_{\rm liq}$ is just given by the 
probability that two NRRW's stay together over a distance $l$, 
\bea
	\label{Pliq1}
	P_{\rm liq}(l=0)&=&\frac{k}{k+1},\\
	\label{Pliq2}
	P_{\rm liq}(l>0)&=&\frac{k-1}{k+1}\frac{1}{k^l}.
\eea	 
Upon injecting (\ref{f(l)}), (\ref{Pliq1}), and (\ref{Pliq2}) in (\ref{Fav2}) 
one may verify that $\<F_d\>_{\rm liq}=0$. 

In the glass phase, $\<P(l)\>_{\rm state}$ is most easily evaluated as 
$N(l)\<\tilde P(l)\>_{\rm state}$, where $N(l)$ is the number of rooted 
NRRW's of length $l$ and $\tilde P(l)$ is the probability for two 
replicas to stay on a specific path of length $l$. 

In the Bethe-Peierls approximation the latter can be computed within 
an enlarged cavity containing all sites of the path. The average over 
the states is done by averaging independently over the local field 
distributions on all neighboring sites, taking into account proper weighting 
factors:
\be
	\<\tilde P(l)\>_{\rm state} = \frac{1}{L}\sum_{a=1}^{L}\frac{\int\!\prod_{\iota \in \mathcal{I}_l}d\rho\left({\bf p}^{(\iota)}\right) \mathcal{P}_{l;a}\, W_{l;{\rm tot}}(\{{\bf p}^{\iota}\}_{\iota \in \mathcal{I}_l})^m}
{\int\!\prod_{\iota \in \mathcal{I}_l}d\rho\left({\bf p}^{(\iota)}\right) W_{l;{\rm tot}}(\{{\bf p}^{\iota}\}_{\iota \in \mathcal{I}_l})^m}\,,
\ee
where we have introduced the set of indices $\mathcal{I}_l$ labeling 
the neighbors of the $l+1$ sites on the path: 
\be
\mathcal{I}_l=\cup_{i_0=1}^{k}\{(0,i_0)\}\cup_{l'=1}^{l-1}\left(\cup_{i_{l'}=1}^{k-1}\{(l',i_{l'})\}\right)\cup_{i_l=1}^{k}\{(l,i_l)\}\,.
\ee
$\mathcal{P}_{l;a}$ denotes the probability, given the local field 
configuration, for two replicas to both stay on the given path up to 
site $l$ and to separate afterwards, under the condition to start off at 
site $0$ with monomer $a$,
\be
\label{Prob_la}
	\mathcal{P}_{l;a}=\frac{\sum^k_{j_2\neq j_1} W^{(j_1)}_{l; a+} W^{(j_2)}_{l; a+}}{W_{l;a}(\{{\bf p}^{\iota}\}_{\iota \in \mathcal{I}_l})^2}\,.
\ee

The weights $ W^{(j)}_{l;a\pm}$ are the Boltzmann factors associated with 
a polymer starting with monomer $a$ on site $0$, staying on the path, and 
leaving it at the site $l$ via neighbor $(l,j)$,
\be 
	W^{(j)}_{l;a\pm} = e^{\beta \mu(l+1) }\left(\sum_{j'=1}^{k}\frac{p^{(0,j')}_{\da (a-1) }}{\psi^{(0,j')}_a} 
\prod_{i=1}^{k}\psi_{a}^{(0,i)}\right)
\prod_{l'=1}^{l-1} \left(\prod_{i=1}^{k-1}\psi_{a\pm l'}^{(l',i)} \right) \left( \frac{p^{(l,j)}_{\ua a\pm (l+1)}}{\psi_{a\pm l}^{(l,j)}} \prod_{i=1}^{k}\psi_{a\pm l}^{(l,i)} \right)\,,
\ee
the sign $\pm$ indicating that monomer indices increase/decrease 
along the path. Notice that in Eq. (\ref{Prob_la}) 
we selected arbitrarily one of the two equivalent directions.
In the above formul\ae, $W_{l;{\rm tot}}$ and $W_{l;a}$ are the 
Boltzmann factors associated with the ensemble of all possible 
configurations on the path, and of the configurations restricted to 
have a monomer $a$ on site $0$, respectively. They are conveniently 
calculated recursively via

\be
\label{recurs}
W_{l;a/{\rm tot}}(\{{\bf p}^{\iota}\}_{\iota \in \mathcal{I}_l}) =C\left({\bf p}^{(l,1)},\dots,{\bf p}^{(l,k)}\right) W_{l-1;a/{\rm tot}}\left(\{{\bf p}^{\iota}\}_{\iota \in \mathcal{I}_{l-1}}| {\bf p}^{(l-1,k)}=I\left({\bf p}^{(l,1)},\dots,{\bf p}^{(l,k)}\right)\right) 
\,,
\ee
where $I$ denotes the cavity iteration functional as defined by (\ref{cavity0}-\ref{cavity2}), and $C$ is the corresponding normalization constant.
The initial conditions for (\ref{recurs}) are simply
\be 
	W_{0;a}\left({\bf p}^{(0,1)},\dots,{\bf p}^{(0,k+1)}\right)=e^{\beta\mu}\sum_{i_1\neq i_2}^{k+1}p^{(0,i_1)}_{1a-1\da}p^{(0,i_2)}_{1a+1\ua}\prod_{j\neq i_1,i_2}^{k+1}\psi^{(0,j)}_{\sigma_a}
\ee
and 
\be 
	W_{0;{\rm tot}}\left({\bf p}^{(0,1)},\dots,{\bf p}^{(0,k)}\right)=w_s\left({\bf p}^{(0,1)},\dots,{\bf p}^{(0,k)}\right).
\ee

%
%
\section{Monte Carlo simulations on the Bethe lattice}
\label{app:mcB}
In this Appendix we describe our approach to numerical simulations
of heteropolymers on the Bethe lattice. In the first part we 
define a model for finite length polymers. In the second one
we present our Monte Carlo algorithm.
%
%
\subsection{Finite length polymers}

We consider a modified ensemble for a varying number of finite length 
random walks. More precisely, a configuration is defined by $n$
mutually-avoiding SAW's. The chain $i$ shall contain $N_i$ monomers,
with $N_1+\dots+N_n =N$. The Hamiltonian (\ref{Hamiltonian})
receives contributions both from self-contacts within a single chain and from
mutual contacts between different chains.
The grand-canonical free energy is
\begin{eqnarray}
-\beta\, \omega_{L}(\beta,\mu,\mu_{\rm end}) = \lim_{V\to\infty}
\frac{1}{V}{\mathbb E}_{\cal G}\log 
\left(\sum_{n\ge 0}\ex^{\beta\mu_{\rm end} n}\sum_{N\ge 0} \ex^{\beta\mu N}
\sum_{\uo}\ex^{-\beta H_{N}(\uo)}\right)\, .
\label{VariableLength}
\end{eqnarray}
We introduced the chemical potential $\mu_{\rm end}$ which couples 
to the number of chain ends in the solution (or, equivalently,
to the number of polymers). The single-polymer ensemble is recovered 
in the $\mu_{\rm end}\to -\infty$ limit.

Extending the cavity formalism to the finite-$\mu_{\rm end}$ case is quite
straightforward. As an illustration, we can easily write down the 
generalization of Eqs. (\ref{cavity0})-(\ref{cavity2}):
\begin{eqnarray}
p_0^{(0)} &=&C^{-1}\prod_{i=1}^{k} \psi_0^{(i)},\\
p_{\ua a}^{(0)} &=&C^{-1}e^{\beta \mu}\prod_{j=1}^k \psi_a^{(j)}
\left\{e^{\beta\mu_{\rm end}}+
\sum_{i=1}^{k} \hat{p}_{\ua a+ 1}^{(i)}\right\},\\
p_{\da a}^{(0)} & = & C^{-1}e^{\beta \mu}\prod_{i=1}^{k} \psi_a^{(i)}
\left\{e^{\beta\mu_{\rm end}}+
\sum_{i=1}^{k} \hat{p}_{\da a-1}^{(i)}\right\},\\
p_{2a}^{(0)} &=& C^{-1}e^{\beta \mu}\prod_{i=1}^{k} \psi_a^{(i)}\left\{
e^{2\beta\mu_{\rm end}}+
e^{\beta\mu_{\rm end}}\sum_{i=1}^{k}(\hat{p}_{\da a-1}^{(i_1)} +
\hat{p}_{\ua a+1}^{(i_2)})+
\sum_{i_1\ne i_2}^k \hat{p}_{\da a-1}^{(i_1)} 
\hat{p}_{\ua a+1}^{(i_2)} \right\}\, ,
\end{eqnarray}
where we used the shorthands $\hat{p}_{\da a-1}^{(i)} \equiv
p_{\da a-1}^{(i)}/ \psi_a^{(i)}$, $\hat{p}_{\ua a+1}^{(i)} \equiv
p_{\ua a+1}^{(i)}/ \psi_a^{(i)}$.
%
%
\subsection{The Monte Carlo algorithm}

\begin{figure}
\centerline{
\epsfig{figure=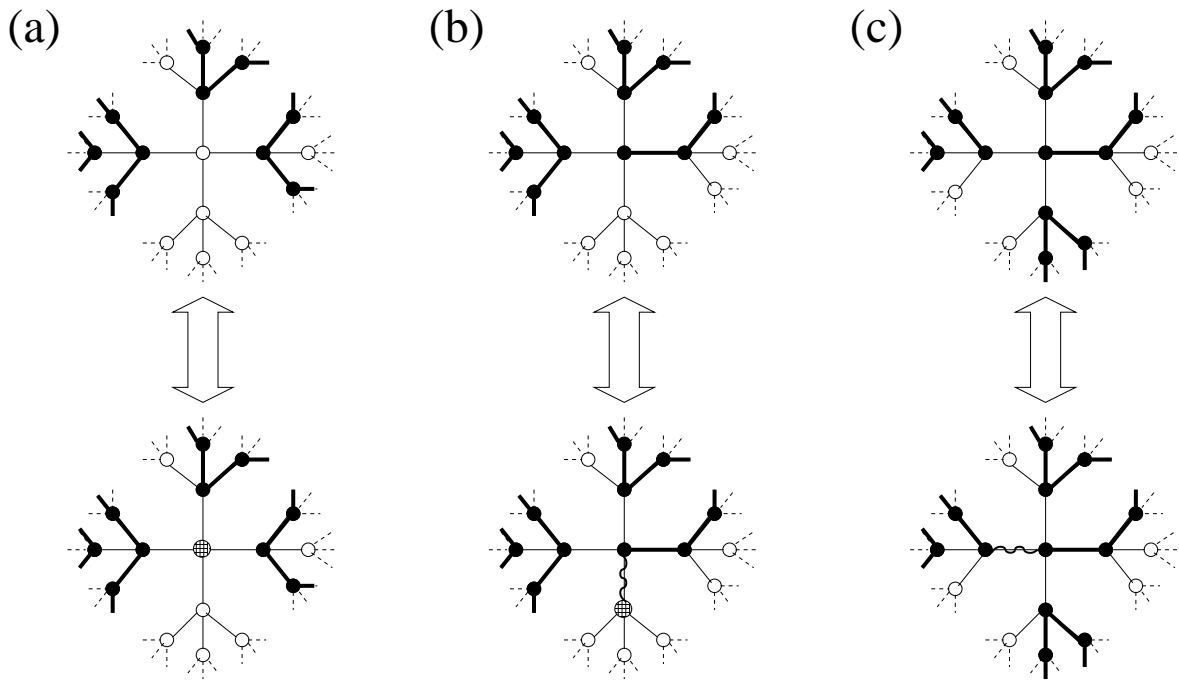,width=0.95\linewidth}}
\caption{The three moves used in our Monte Carlo simulations 
on the Bethe lattice. The monomers (chain links) which change because
of the move are represented with hatched circles (wiggly lines).}
\label{BetheMoves}
\end{figure}
As already mentioned in Sec. \ref{NumericalBethe}, numerical simulations
of long fixed-length polymers are quite difficult on the Bethe lattice.
We thus resort to simulating the variable-length ensemble 
corresponding to the free-energy (\ref{VariableLength}). The 
algorithm includes three types of moves illustrated graphically 
in Fig. \ref{BetheMoves}: {\bf (a)} monomer insertion/deletion;
{\bf (b)} chain extension/reduction; {\bf (c)} two chain junction/disjunction.
It is straightforward to show that these three moves ensure ergodicity.

At each step of the algorithm the type of move 
and the location in the graph are chosen randomly. The move
is then accepted according to the Metropolis rule in such a way as to
satisfy detailed balance with respect to the variable length
ensemble  (\ref{VariableLength}). Evidently the algorithm is
more efficient for moderate lengths of the polymers, i.e., 
not too large values of $|\mu_{\rm end}|$. It can be therefore 
convenient, for producing equilibrated configurations,
to gradually decrease $|\mu_{\rm end}|$ to the desired value.
%
%
\section{Numerical solution of the 1RSB cavity equations with 
population dynamics}
\label{app:popdyn}
The cavity recursion equation in the form (\ref{onestepcavity}) suggests 
a numerical solution by an iterative population dynamics~\cite{cavityT}: 
The distribution of local fields 
$\rho({\bf p})$ is represented by a (finite) population of fields. An 
iteration step in the dynamics consists in choosing at random $k$ 
``parent'' members ${\bf p}^{(i)}$ of the population and calculating 
the corresponding cavity field ${\bf p}^{(0)}=I(\{{\bf p}^{(i)}\})$ from 
(\ref{cavity0})-(\ref{cavity2}). This new field is then exchanged against 
an old field in the population with probability  
${C[\{p^{(i)}\}]^m}/C_{\rm max}^m$, proportional to the reweighting 
$C[\{p^{(i)}\}]^m$ (normalized so as to make sure that the probability 
never exceeds 1). If the dynamics converges to a stationary distribution, 
its density satisfies the recursion equation (\ref{onestepcavity}).

In the soft glass phase, the iteration converges rapidly since the 
distribution of fields remains centered around the unstable liquid fixed 
point. However, the algorithm considerably slows down in the frozen phase 
where the fields have strong biases towards given conformations. Since the 
biases of the $k$ parent members are only rarely compatible with each other, 
the reweighting is usually very small. The population dynamics is then 
dominated by rare events with a low degree of frustration. Obviously, 
the probability of frustrated events rapidly increases with the number 
of different local conformations and thus with the length of the period 
$L$. For this reason we have limited our numerical simulations in the 
frozen glass phase to populations of 4000 fields for chains with $L=20$.

\bibliographystyle{plain}
\bibliography{../Bib/references}

\end{document}